\documentclass[aps,pre,twocolumn,superscriptaddress,nofootinbib]{revtex4}

\usepackage{graphicx}
\usepackage{amssymb}
\usepackage{amsmath}
\usepackage{amsfonts}
\usepackage{color}
\usepackage{hyperref}

\DeclareMathAccent{\ring}{\mathalpha}{operators}{"17}
\providecommand{\st}[1]{_{\text{#1}}}

\providecommand{\ut}[1]{^{\text{#1}}}
\providecommand{\ten}[1]{\bv{#1}}

\def\onehalf{\frac{1}{2}}

\def\bra{\ensuremath{\langle}}
\def\ket{\ensuremath{\rangle}}

\def\ueq{\ut{eq}}

\def\pd{\partial}

\def\im{\mathrm{i}}

\def\cslb{\sigma_s}
\def\kv{\bv{k}}
\def\qv{\bv{q}}
\def\uv{\bv{u}}

\def\cv{\bv{c}}

\def\rv{\bv{r}}
\def\b0{\bv{0}}

\def\ra{\rightarrow}
\def\Fcal{\mathcal{F}}
\def\Dcal{\mathcal{D}}
\def\Ocal{\mathcal{O}}
\def\Pt{\ten{P}}

\newcommand{\bitem}{\begin{itemize}}
\newcommand{\eitem}{\end{itemize}}
\newcommand{\benum}{\begin{enumerate}}
\newcommand{\eenum}{\end{enumerate}}
\newcommand{\bblock}[1]{\begin{block}{#1}}
\newcommand{\eblock}{\end{block}}
\newcommand{\bmini}[1]{\begin{minipage}{#1}}
\newcommand{\emini}{\end{minipage}}
\newcommand{\btab}[1]{\begin{tabular}{#1}}
\newcommand{\etab}{\end{tabular}}
\newcommand{\btabn}[1]{\begin{tabular}{#1}}
\newcommand{\etabn}{\end{tabular}}
\newcommand{\beq}{\begin{equation}}
\newcommand{\eeq}{\end{equation}}
\newcommand{\beqn}{\begin{equation*}}
\newcommand{\eeqn}{\end{equation*}}
\newcommand{\bmult}{\begin{multline}}
\newcommand{\emult}{\end{multline}}
\newcommand{\bsplit}{\begin{split}}
\newcommand{\esplit}{\end{split}}

\newcommand{\bv}[1]{\mathbf{#1}}

\begin{document}
 \title{Simulation of static critical phenomena in non-ideal fluids \\with the Lattice Boltzmann method}
 \author{M. Gross}
 \email{markus.gross@rub.de}
 \affiliation{Interdisciplinary Centre for Advanced Materials Simulation (ICAMS), Ruhr-Universit\"at Bochum, Stiepeler Strasse 129, 44801 Bochum, Germany}
 \author{F. Varnik}
 \affiliation{Interdisciplinary Centre for Advanced Materials Simulation (ICAMS), Ruhr-Universit\"at Bochum, Stiepeler Strasse 129, 44801 Bochum, Germany}
 \affiliation{Max-Planck Institut f\"ur Eisenforschung, Max-Planck Str.~1, 40237 D\"usseldorf, Germany}

\begin{abstract}
A fluctuating non-ideal fluid at its critical point is simulated with the Lattice Boltzmann method.
It is demonstrated that the method, employing a Ginzburg-Landau free energy functional, correctly reproduces the static critical behavior associated with the Ising universality class.
A finite-size scaling analysis is applied to determine the critical exponents related to the order parameter, compressibility and specific heat. A particular focus is put on finite-size effects and issues related to the global conservation of the order parameter.
\end{abstract}


\maketitle

\section{Introduction}
The theory of critical phase transitions has received considerable attention and seen remarkable progress in the past decades \cite{amit_book, yeomans_book, pelissetto_vicari_review2002, hohenberg_halperin, folk_moser_review2006}.
The importance of critical phenomena stems from the fact that systems, whose microscopic behavior can be very diverse, nevertheless share the same universal properties close to their critical points and thus belong to the same universality class.
Universality classes are defined only by a few characteristic properties, such as the number of components of the order parameter, the dimensionality of space, the couplings to other dynamical quantities in the system and the presence of conservation laws.
It is important to realize that universality classes are different regarding static and dynamic critical properties: While, for instance, the uniaxial ferromagnet and a pure fluid both show Ising-type static critical behavior, their critical dynamics is decisively different \cite{hohenberg_halperin}.

Most critical properties for standard bulk systems, such as critical exponents and amplitude ratios, are nowadays known with high precision due to the combined effort of experimental, theoretical and simulation approaches.
Thus, in recent years, the focus has moved on to the study of critical phenomena in more complex situations, such as under non-equilibrium conditions \cite{calabrese_aging_jphysa2005}, at surfaces \cite{pleimling_surf_review2004, gambassi_thinfilms_jstatp2006, parry_critical_wetting_jlowt2009} or in complex fluids \cite{royall_natphys2007}.
Here, it is hoped that peculiar fluctuation induced effects, such as the critical Casimir effect \cite{krech_casimir_book, hertlein_nature2008}, can be utilized for novel applications.
Due to the increasing complexity of such systems, simulation approaches to critical dynamics in fluid systems thus become an indispensable tool.

While dynamic critical phenomena of fluids have been extensively studied theoretically and by experiment \cite{hohenberg_halperin, sengers_supercritical_1994, folk_moser_review2006}, their simulation has only been recently approached via Molecular Dynamics \cite{hamanaka_onuki_pre2005, chen_dynamic_prl2005, jagannathan_prl2004,  das_binary_prl2006, das_binary_jcp2006, roy_das_epl2011}. However, system sizes are rather limited and certain transport coefficients, such as the shear viscosity, are notoriously hard to determine with sufficient accuracy.
Recently, the Lattice Boltzmann (LB) method -- being a well-established and efficient solver of the Navier-Stokes equations -- has been extended to deal with thermal fluctuations in liquid-vapor systems \cite{gross_flb_2010} as well as binary fluids \cite{sumesh_binary_2011}.
Thus, the LB method appears to be a promising candidate for the simulation of critical phenomena in simple and complex fluids.

As a first step toward this aim, the current work presents simulation results on the \textit{static} critical behavior of a non-ideal fluid obtained with the fluctuating LB model introduced in \cite{gross_flb_2010}.
Dynamic critical properties will discussed in a separate paper \cite{gross_critical_dynamics}.
In the present model, the fluctuating hydrodynamic equations for the density and momentum of an isothermal, non-ideal fluid are solved via a Langevin approach.
All simulations are performed in two dimensions, which -- besides computational efficiency -- has the advantage that critical properties can be much easier assessed than in 3D, as fluctuation effects are generally more pronounced in lower dimensions.
Since the model is governed by a one-component Ginzburg-Landau $\phi^4$-free energy functional, the static critical properties are expected to be described by the 2D Ising universality class \cite{wilson_kogut_1974, leguillou_zinnjustin_prb1980, zinnjustin_qftcrit_book, binney_book}.
This prediction is indeed borne out by the present LB simulations, which are also in line with previous Monte-Carlo investigations of the two-dimensional $\phi^4$-model \cite{cooper_freedman_nuclpb1982, callaway_nuclpb1984, bruce_jphysa1985, milchev_1986, toral_prb1990, mehlig_zphysb1992, loinaz_prd1998, marrero_physlettb1999, de_prd2005}.
A crucial issue in a hydrodynamics based simulation approach is the global conservation of the order parameter (here, the density), which complicates the application of the finite-size scaling technique used to extract critical properties in a finite system \cite{rovere_jpcm1990, rovere_zphysb1993}.
The aim of the present work is to provide a thorough assessment of the LB method in the critical fluctuation regime and demonstrate that, despite the above mentioned complications, the method is able to successfully simulate critical fluctuations in fluids. At the same time, important issues that might be useful for further applications and extensions of the method shall be highlighted.
The paper is written in a self-contained manner and is hoped to provide also a researcher unacquainted with critical phenomena with sufficient background information.

The outline of the paper is as follows. In section II, the critical properties of the Ginzburg-Landau theory are reviewed. In particular, the effects of a finite system size and the global conservation of the order parameter are discussed. Also, the order-parameter distribution as a fundamental quantity to extract information on critical as well as non-critical properties is introduced.
Section III discusses the simulation method and contains a number of remarks on the correct choice of simulation parameters.
In section IV, simulation results on the structure factor and important thermodynamic quantities are presented and compared to theoretical predictions and previous works.

\section{Theory}
\subsection{Ginzburg-Landau model}
\subsubsection{Introduction}
\label{sec:gl}
As the present simulation approach to critical phase transitions is based on fluctuating hydrodynamics (see section \ref{sec:method}), the fundamental quantity for our purpose is the density field of the fluid, $\rho(\rv)$. From the density, an \emph{order parameter} $\phi(\rv)$ can be defined as
\beq \phi(\rv)= \frac{\rho(\rv)-\rho_0}{\rho_0}\,, \label{op}\eeq
where the reference density $\rho_0$ is taken as the global average, $\rho_0=\int d\rv \rho(\rv)/V$, with $V$ being the system volume.
The equilibrium behavior of the order parameter is governed by a Ginzburg-Landau free energy functional
\beq
\Fcal[\phi] = \int d\rv \left[\frac{\kappa}{2} |\nabla\phi|^2 + f_0(\phi) - h\phi\right]\,,
\label{fef}
\eeq
with $f_0$ being a Landau potential
\beq f_0(\phi) = \frac{r}{2} \phi^2 + \frac{u}{4}\phi^4\,,
\label{landau-pot}
\eeq
and $h$ an external field which is used to define response functions.
As usual $\kappa$ and $u$ are strictly positive, while the coefficient $r$ of the quadratic term can be either positive or negative, leading to either a single minimum or a double-well form of the Landau free energy.
Thermal fluctuations lead to a equilibrium distribution of the order parameter according to the probability density
\beq P[\phi] = \frac{1}{Z} e^{-\Fcal[\phi]/k_B T}\,.
\label{prob-dist}
\eeq
Here, $k_B$ is the Boltzmann constant and $T$ is the temperature. Note that no dependence of the coefficient $r$ in the Landau potential on the temperature $T$ is assumed. Rather, $r$ is considered as an independent quantity representing the appropriate temperature measure in the context of the Ginzburg-Landau model (see next section).
The partition sum $Z$ is given by
\beq Z = \int \Dcal \phi e^{-\Fcal[\phi] / k_B T}\,,
\label{part-sum}
\eeq
where $\int \Dcal\phi$ denotes the integration over all possible realizations of the order parameter distribution. On a $d$-dimensional lattice of volume $V$ with a total of $N$ lattice points, the order parameter $\phi$ is specified by its $N$ values $\phi_i\equiv \phi(\rv_i)$ and the functional integral is regularized as $\int\Dcal\phi \rightarrow \Pi_{i=1}^N \int d\phi_i$. Discrete equivalents for the derivative operators can be found in \cite{gross_flb_2010}.
The partition sum \eqref{part-sum} makes it possible to define a thermodynamic Helmholtz free energy $F$ and a corresponding density $f$ in the usual way as
\beq F = f V = -k_B T \log Z\,.
\label{HelmholtzF}
\eeq

From the free energy, eq.~\eqref{HelmholtzF}, a global, ensemble-averaged order parameter $M$ and an associated susceptibility $\chi$ can be formally defined as response functions with regard to the external field:
\begin{align}
M &= -\frac{\partial f}{\partial h} = \frac{1}{V}\int d\rv \bra \phi(\rv) \ket = \bra m\ket \label{op-cont}\,,\\
\begin{split}
\chi &= -\frac{\partial^2 f}{\partial h^2} = \frac{1}{k_B T V}\int d\rv d\rv' \big[\bra \phi(\rv)\phi(\rv')\ket -\bra \phi(\rv)\ket\bra \phi(\rv')\ket \big] \\&= \frac{V}{k_B T}\left(\bra m^2\ket - \bra m\ket^2\right)\,,
\end{split}
\label{chi-cont}
\end{align}
where $m \equiv \int d\rv \phi/V$ and the brackets denote average with respect to the distribution $P$, that is, $\bra g(\phi)\ket \equiv \frac{1}{Z}\int \Dcal\phi g(\phi) P[\phi]$ for an arbitrary function $g$ of $\phi$.
A further quantity of interest is the spatial correlation function of the order-parameter fluctuations (structure factor)
\beq C(\rv)\delta(\rv-\rv')=\bra (\phi(\rv)-\bra\phi\ket)(\phi(\rv')-\bra\phi\ket)\ket \,,
\eeq
and its Fourier transform $C(\kv)$. Note that translational invariance is assumed in the above equation.
Related to the correlation function is a non-local susceptibility $\chi(\rv)=C(\rv)/k_B T$, which can be defined analogously to eq.~\eqref{chi-cont} via the linear response to a spatially dependent external field.
The definition of the specific heat, which quantifies the thermal response, requires some care, since a temperature change can be effected in several ways, depending on the parameterization of the model. Here, the field theoretic convention \cite{amit_book, zinnjustin_qftcrit_book} is followed and the specific heat is defined as the response with respect to a change of the coefficient $r$,
\begin{multline}
 c_H=\frac{\partial^2 f}{\partial r^2} = \frac{1}{4 k_B T V}\int d\rv d\rv' \big[\bra \phi^2(\rv)\phi^2(\rv')\ket \\-\bra \phi^2(\rv)\ket\bra \phi^2(\rv')\ket \big] = \frac{V}{k_B T} \left(\bra E^2\ket -\bra E\ket^2 \right)\,,
\label{spec-heat-cont}
\end{multline}
where $E\equiv \int d\rv \phi^2/2V$ represents the most singular part of the local energy $\Fcal$.

It is often convenient to rewrite the Ginzburg-Landau free energy in terms of a minimal number of parameters. To this end, we note first that the temperature only appears as an overall scale factor in the Boltzmann weight, eq.~\eqref{prob-dist}, and can thus be absorbed in the definition of the coupling constants. Second, the coefficient of the square-gradient term can be fixed to $1/2$ by rescaling the order parameter field as $\phi = \tilde\phi/\sqrt{\kappa/ k_B T}$. The reparameterized free energy functional reads
\beq \tilde \Fcal[\tilde\phi] =\Fcal[\phi]/k_B T = \int d\rv \left(\onehalf |\nabla\tilde\phi|^2 + \frac{\tilde r}{2}\tilde \phi^2 + \frac{\tilde u}{4}\tilde \phi^4 - \tilde h \tilde \phi \right)
\label{fef-rescaled}
\eeq
where
\beq \tilde r = \frac{r}{\kappa}\,,\quad \tilde u = \frac{u k_B T}{\kappa^2}
\label{reduced-param}
\eeq
are the two remaining independent coupling constants. Correspondingly, the Boltzmann weight in eq.~\eqref{prob-dist} becomes $e^{-\tilde \Fcal[\tilde\phi]}$.
The functional \eqref{fef-rescaled} is the usual starting point for field-theoretic studies of the Ginzburg-Landau model \cite{amit_book, binney_book, zinnjustin_qftcrit_book}.
In the following, both parameterizations, eqs.~\eqref{fef} and \eqref{fef-rescaled}, of the model shall be used (dropping the tilde on $\tilde\phi$ and $\tilde h$ for readability).

\subsubsection{Critical behavior}
\label{sec:crit_behavior}
Theoretically, the Ginzburg-Landau free energy functional of eq.~\eqref{fef} can be obtained as a coarse-grained description of some microscopic degrees of freedom, for example, spins on a lattice or molecules of a fluid. Accordingly, the order parameter is defined as an average over some coarse-graining length. In this regard, the free energy functional $\Fcal$ can be considered as an effective Hamiltonian from which a partition function and a corresponding Helmholtz free energy can be obtained.
Close to the upper critical dimension $d_c=4$ of the Ginzburg-Landau model, standard renormalization group arguments show that all terms of higher order than the $\phi^4$-term are irrelevant at the critical point and the simple Landau potential of eq.~\eqref{landau-pot} indeed describes the universal critical properties of all systems with the same symmetry property of the order parameter \cite{wilson_kogut_1974, zinnjustin_qftcrit_book, binney_book, amit_book}.
Monte-Carlo simulations \cite{cooper_freedman_nuclpb1982, callaway_nuclpb1984, bruce_jphysa1985, milchev_1986, toral_prb1990, mehlig_zphysb1992, loinaz_prd1998, marrero_physlettb1999, de_prd2005} as well as theoretical arguments invoking conformal invariance \cite{zamolodchikov_conformal_1986, cardy_leshouches_1988, morris_physlettb1995} have confirmed that the \textit{two-dimensional} $\phi^4$-model with a scalar order parameter belongs to the 2D Ising universality class.

In the context of liquid-vapor criticality, a few remarks on the applicability of the Ginzburg-Landau model to a real critical fluid are in order:
In general, the coexistence curve of a real fluid is not symmetric; instead the liquid and vapor densities follow a relation which, in its simplest form, is known as the ``law of rectilinear diameter'' (see, e.g., \cite{greer_moldover_anomalies_rev1981}). Such asymmetry is, for instance, predicted by the van der Waals equation of state \cite{wyczalkowska_sengers_crit_vdw_2004} and could be accounted for in a Ginzburg-Landau scheme by adding a term $\propto \phi^5$ to the Landau potential \cite{nicoll_quintic_pra1981}.
The asymmetry is caused by the fact that the relevant ordering- and thermal scaling fields, $h$ and $r$ [eq.~\eqref{landau-pot}], which are characteristic for an Ising-like system possessing a ``particle-hole'' symmetry, are linear combinations of the physical variables temperature and chemical potential. Similarly, the order parameter $\phi$ and the energy density (which, in the present case, is not an independent field but related to $\phi^2$) are linear combinations of the physical mass and energy density \cite{mermin_rehr_prl1971, pokrovski_1973, onuki_book}. Recently, it has been shown that also the mixing of the pressure into the scaling fields is important in order to account for certain critical anomalies \cite{fisher_orkoulas_prl2000, kim_fisher_pre2003, anisimov_wang_prl2006, wang_anisimov_pre2007}. The critical behavior of real fluids is thus described by the Ising-universality class in the sense of a mapping relation between physical and Ising variables.

Neglecting thermal fluctuations and evaluating the partition sum \eqref{part-sum} only along its saddle-point, defines the \textit{mean-field approximation}, for which the critical point occurs for $\tilde r = r=0$.
However, when the Landau potential becomes very shallow, thermal fluctuations can significantly contribute to the functional integral in \eqref{part-sum}, leading eventually to a breakdown of mean-field theory.
The critical point of the full Ginzburg-Landau model in fact occurs at a slightly negative $\tilde r$, which, due to the non-linear interactions between the fluctuations, depends on the non-linear coupling $\tilde u$ \cite{chaikin_book} (see below).
The ``distance'' to the critical point $\tilde r_c$ can be defined in terms of a reduced dimensionless temperature
\beq \theta\equiv \frac{\tilde r_c-\tilde r}{\tilde r_c} = \frac{r_c-r}{r_c}\,,
\label{t-red}
\eeq
where fixed $\kappa$ and $T$ are assumed in the last equation.
The above definition ensures that $\theta>0$ in the disordered phase (super-critical regime) and $\theta<0$ in the ordered phase (sub-critical regime). In mean-field theory, $\tilde r_c = r_c=0$; thus, definition \eqref{t-red} must be replaced by $\theta=r/a$, where $a$ is a suitable constant in order to make $\theta$ dimensionless.

\begin{table*}[t]
\begin{center}
\begin{tabular}{c | c c c c c}
\hline\hline
Exponent & $\alpha$ & $\beta$ & $\gamma$ & $\nu$ & $\eta$ \\
Quantity & specific heat & order parameter & susceptibility & correlation length & structure factor \\
Definition & $c_H\propto \theta^{-\alpha}$ & $M \propto \theta^\beta$ & $\chi\propto \theta^{-\gamma}$ & $\xi\propto \theta^{-\nu}$ & $C(k)\propto k^{-2+\eta}$ \\
\hline
Mean-field & 0 (disc.) & 1/2 & 1 & 1/2 & 0 \\
Ising 2D & 0 (log.) & 1/8 & 7/4 & 1 & 1/4\\
\hline\hline
\end{tabular}
\end{center}
\caption{Critical exponents for the mean-field and the 2D-Ising universality class. $\theta$ is the reduced temperature. The specific heat is discontinuous in mean-field theory and logarithmically divergent in the 2D Ising case.
}
\label{tab:exp}
\end{table*}

Close the critical point, thermodynamic quantities typically show a power-law dependence on the reduced temperature $\theta$, with exponents that are identical for all systems within the same universality class \cite{chaikin_book, yeomans_book}.
Two-scale factor universality implies that the singular dependence of the Helmholtz free energy, eq.~\eqref{HelmholtzF}, on the two relevant scaling variables temperature $\theta$ and external field $h$ is given by
\beq f\st{sing}(\theta,h) = |\theta|^{2-\alpha} f_\pm(h/|\theta|^{\beta\delta})\,,
\label{F-sing}
\eeq
where $f_\pm$ is a universal scaling function (up to metrical factors) and $\alpha$ and $\delta$ are critical exponents.
From the above relation, the critical behavior of the order parameter, susceptibility and specific heat follows as
\beq\begin{aligned}
M &  \simeq B (-\theta)^\beta\qquad (\theta<0)\,,\\
\chi &  \simeq \Gamma_\pm |\theta|^{-\gamma}\,,\\
c_H & \simeq  A_\pm |\theta|^{-\alpha} \overset{\text{2D}}{\longrightarrow} A_\pm\log(\theta)\,,
\end{aligned}
\label{powerlaws}
\eeq
where $\Gamma_\pm$, $B$ and $A_\pm$ are non-universal amplitudes ($\pm$ refers to whether the critical point is approached from above or below). In the two dimensional Ginzburg-Landau model, the specific heat has a logarithmic divergence (which is conventionally indicated by an exponent $\alpha=0$) \footnote{For other parameterizations of the temperature dependence of the model, there can also be a regular contribution to the specific heat.}.
The correlation length $\xi$ diverges as
\beq \xi\propto \theta^{-\nu}\,,\eeq
while the correlation function at criticality assumes a power law,
\beq C_\text{crit}(\kv) \propto k^{-2+\eta}\,,
\label{C-crit}
\eeq
expected to be valid for $k\gtrsim 1/\xi$ \cite{fisher_burford_pr1967, tracy_mccoy_prb1975, wu_mccoy_prb1976}.
The values of the critical exponents are collected in Table~\ref{tab:exp}.

\begin{figure}[b]
\centering
    \includegraphics[width=0.75\linewidth]{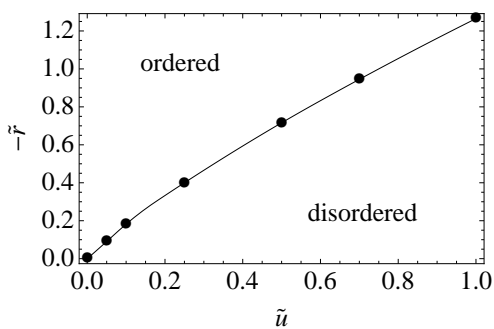}
   \caption{Phase diagram for the $\phi^4$-model on a square lattice. Data (symbols) are taken from \cite{toral_prb1990}. The critical line continues to the Ising-limit, $\tilde u\rightarrow \infty$, $\tilde r\rightarrow -\infty$.}
    \label{fig:phase-diag}
\end{figure}
From the reparameterized free energy, eq.~\eqref{fef-rescaled}, we see that, in contrast to the Ising model, where only one coupling constant and thus a single critical point exists, the Ginzburg-Landau model entails a line $\tilde r_c(\tilde u)$ of critical points \cite{chaikin_book}. The universal critical properties of all points on the critical line are controlled by the renormalization group fixed point, which is expected to belong to the Ising universality class.
The critical line of the Ginzburg-Landau model on a square lattice has been obtained in previous works via Monte-Carlo simulations \cite{milchev_1986, toral_prb1990, mehlig_zphysb1992, loinaz_prd1998, de_prd2005}. Figure \ref{fig:phase-diag} shows the corresponding phase diagram taken from \cite{toral_prb1990}. Note that the critical value for $\tilde r$ decreases with increasing interaction strength $\tilde u$.

Two particular limits on the critical line deserve further remarks \cite{bruce_advphys1980}: In the ``order-disorder limit'' (Ising limit), which is reached for $\tilde r\rightarrow -\infty$, $\tilde u\rightarrow \infty$ with $\tilde r/\tilde u=\text{const.}$, the potential has two minima separated by an infinitely high barrier and the lattice free energy functional becomes formally identical to the Ising Hamiltonian.
On the other hand, the case $\tilde r\rightarrow 0$, $\tilde u\rightarrow 0$ defines the so-called ``displacive limit'', where the free energy functional is dominated by the gradient-term interaction and the central potential barrier is low compared to the thermal energy. This limit is particularly important in the case of structural phase transitions \cite{bruce_advphys1980}.
For an infinite system, critical properties of all points on the critical line are universal and of Ising type, except in the displacive limit, where Gaussian critical behavior is expected. For finite systems, close proximity to the displacive limit can lead to an undesired masking of Ising-type critical behavior \cite{bruce_advphys1980, milchev_1986, amit_book}.

In the parametrization \eqref{fef-rescaled} of the Ginzburg-Landau model, dimensional analysis shows that the length dimension of $\tilde \phi$ is
$[\tilde\phi] = L^{1-d/2}$, which immediately fixes the dimensions of the coupling constants as $[\tilde r] = L^{-2}$ and $[\tilde u] = L^{d-4}$.
Thus, a dimensionless coupling constant can be defined as
\beq \lambda \equiv \frac{\tilde u}{\tilde r^{(4-d)/2}}\,.
\label{diml_coupl_gen}
\eeq
In 2D, $\tilde r$ and $\tilde u$ have the same dimensions and definition \eqref{diml_coupl_gen} becomes particularly simple:
\beq \lambda =\frac{\tilde u}{\tilde r} = \frac{u\, k_B T}{r \kappa}\,.
\label{diml_coupl}
\eeq

Below, some important analytical approximations to the Ginzburg-Landau model, which will be useful in analyzing the simulation results, is recapitulated briefly.

\subsubsection{Mean-field theory}
\label{sec:mean-field}

In the mean-field approximation, fluctuations around the order-parameter distribution $\bar\phi$ that globally minimizes the Ginzburg-Landau functional are neglected \cite{chaikin_book, yeomans_book}.
This approximation underlies most non-ideal fluid LB models without thermal fluctuations and has been studied extensively in this context (see, e.g., \cite{martys_critical_pre2001, kikkinides_consistency_2008}).
The mean-field free energy is given by $F_0=\Fcal(\bar\phi)$ and admits for two fundamental equilibrium solutions. One corresponds to a spatially uniform value of the order parameter given by
\beq \bar \phi = \left\{\begin{aligned}0 \qquad &(r\geq 0)\\ \pm \sqrt{-r/u} \qquad & (r< 0) \end{aligned}\right.
\label{phi0}
\eeq
and an associated mean-field susceptibility,
\beq \chi = \left\{\begin{aligned}r^{-1}\qquad &(r> 0) \\ (-2r)^{-1}\qquad &(r<0)\,.\end{aligned}\right.
\label{chiMF}
\eeq
In this case, the mean-field free energy amounts to $F_0= V f_0(\bar\phi)$.
In addition, for $r<0$, there exists a solution describing an interface between the two free energy minima of the form
\beq \bar \phi(z) = \sqrt{-r/u}\tanh\left(z/w\right)
\label{intprof}
\eeq
with
\beq w=(-2\kappa/r)^{1/2}
\label{intw}
\eeq
being the interface width. Note that, for the present definition of the interface profile, $w$ is related to the mean-field correlation length $\xi$ [eq.~\eqref{correlMF}] by $w=2\xi$.
The surface tension associated with the planar interface solution \eqref{intprof} is given by
\beq \sigma = \frac{2}{3}\sqrt{-\frac{2\kappa r^3}{u^2}}\,.
\label{surften}
\eeq

\subsubsection{Fluctuations}
\label{sec:fluct}

Thermal fluctuations around a uniform state can be systematically studied by splitting the order parameter into a uniform mean-field part and a spatially inhomogeneous part $\phi(\rv)=\bar \phi+\delta\phi(\rv)$, where $\bar\phi=\bra \phi\ket$ represents the average order parameter.
Expanding $\Fcal$ in the fluctuations $\delta \phi$ and treating the quartic anharmonicity as a perturbation makes it possible to compute the correlation function as a series of Gaussian averages, which can be conveniently represented in terms of Feynman diagrams. To zeroth order in the non-linear coupling $u$, one obtains the \textit{Ornstein-Zernike} (or \textit{Gaussian}) expression for the correlation function,
\beq C_0(\kv)= \frac{k_B T}{c r+\kappa \kv^2} = \frac{k_B T}{\kappa}\frac{1}{ \xi^{-2}+\kv^2} = \frac{k_B T\chi}{1+\kv^2\xi^2}\,,
\label{CMF}
\eeq
where $c=1$ for $r>0$ and $c=-2$ for $r<0$. In the phase-coexistence regime, the constant $c$ accounts for the leading order contribution of the nonlinear term to the correlation length. It is important to emphasize that, due to the assumption of translational invariance, the above expression for the structure factor for $r<0$ holds only in homogeneous states.
In the above equation,
\beq \xi = \sqrt{\kappa \chi} = \left\{\begin{aligned}\sqrt{\frac{\kappa}{r}}\qquad &(r> 0) \\ \sqrt{\frac{\kappa}{-2r}}\qquad &(r<0)\end{aligned}\right.
\label{correlMF}
\eeq
is the mean-field correlation length and $\chi$ is the mean-field susceptibility, eq.~\eqref{chiMF}, which is related to the correlation function by $\chi=\lim_{\kv\ra 0} C_0(\kv)/k_B T$.

In sec.~\ref{sec:crit_behavior} it was shown that the properties of the fluctuating Ginzburg-Landau model are basically governed by the dimensionless coupling constant $\lambda$ of eq.~\eqref{diml_coupl}.
It is informative to express this constant in terms of the physically more relevant parameters correlation length $\xi$, susceptibility $\chi$, surface tension $\sigma$ and order parameter $\bar \phi$: \beq \lambda \sim -\frac{k_B T \chi}{\bar\phi^2\xi^2} \sim -\frac{k_B T \bar\phi^2}{\sigma^2 \chi}\,. \eeq
This shows that increasing the noise temperature $T$ in the ordered state (small negative $\lambda$) brings one always closer to the critical point, unless, for instance, the surface tension or the density ratio ($\bar\phi$) is increased accordingly.

According to the \textit{Ginzburg criterion}, mean-field theory remains valid as long as the mean amplitude of fluctuations $\delta\phi$ around the average order-parameter value $\bar \phi$ [eq.~\eqref{phi0}] remains much smaller than $\bar \phi$ itself,
\beq \sqrt{\bra \delta \phi^2 \ket} \lesssim \bar \phi\,.
\label{ginzbg_crit}
\eeq
By eq.~\eqref{CMF}, $\bra \delta \phi^2\ket = \bra \delta \phi(\rv) \delta \phi(\bv{0})\ket_{\rv=\bv{0}} = \int d\qv C(\qv) \sim k_B T \chi \xi^{-d}$,
and using $\xi=\sqrt{\kappa/2|r|}$ [eq.~\eqref{correlMF}], $\chi=1/2|r|$ [eq.~\eqref{chiMF}] and $\bar \phi = \sqrt{|r|/u}$, the Ginzburg criterion amounts to (neglecting numerical prefactors)
\beq \frac{k_B T }{ \kappa^{d/2}} \lesssim \frac{r^{2-d/2}}{u}\,.
\eeq
We see that, for $d>4$, the right hand side diverges for small $r$, and thus mean-field theory remains valid near the critical point. In contrast, for $d<4$, the Ginzburg criterion is violated for sufficiently small $r$, indicating a breakdown of mean-field theory.
Comparing definition \eqref{diml_coupl_gen} with the Ginzburg criterion, one finds that fluctuation corrections to mean-field theory become significant for $|\lambda| \gtrsim O(1)$.

\subsubsection{Perturbation theory}
\label{sec:perturb}

\begin{figure}[t]
\centering
    \includegraphics[width=0.75\linewidth]{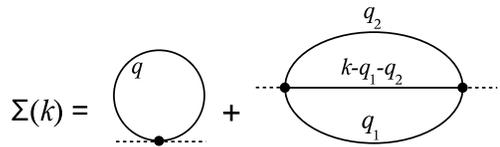}
   \caption{Expansion of the self-energy $\Sigma$ in a self-consistent scheme up to $O(u^2$). Thick lines represent the full correlation function $C$. Dashed lines indicate amputated legs carrying the external wavevector $\kv$.}
    \label{fig:self-en}
\end{figure}

Not too close to the critical point, the effects of fluctuations on observable quantities are -- at least qualitatively -- captured by perturbation theory, which shall be briefly summarized here \cite{binney_book, amit_book,parisi_book, chaikin_book}.
The effect of the non-linear interactions between the fluctuation modes can be captured in terms of a \emph{self-energy} $\Sigma(\kv)$, which is defined by the resummed perturbation expansion (``Dyson equation'') of the full correlation function $C$ as \cite{binney_book, amit_book, parisi_book, zinnjustin_qftcrit_book}
\beq C(\kv) = \frac{1}{C_0^{-1}(\kv) + \Sigma(\kv)}\,.
\label{C_full}
\eeq
$\Sigma$ is given by the sum of all two-point one-particle irreducible diagrams.
In the symmetric phase, the diagrammatic expansion of the self-energy in self-consistent scheme up to second order in the coupling $u$ is shown in Fig.~\ref{fig:self-en}, where the solid lines represent the full correlation function $C$.
The one- and two-loop-contributions to the self-energy are given by \cite{binney_book, amit_book, parisi_book, zinnjustin_qftcrit_book}
\beq \Sigma^{(1)} = \frac{3u}{k_B T}\int^\Lambda \frac{d\qv}{(2\pi)^d} C(\qv)\,,
\label{self1}
\eeq
\begin{multline}
\Sigma^{(2)}(\kv) = -6\left(\frac{u}{k_B T}\right)^2 \int^\Lambda \frac{d\qv_1}{(2\pi)^d}\frac{d\qv_2}{(2\pi)^d} \\ C(\qv_1)C(\qv_2)C(\kv-\qv_1-\qv_2)\,.
\label{self2}
\end{multline}
The notation $\int^\Lambda$ indicates that the integral has to be cut off at a wavenumber $\Lambda$. In the present case, the cut-off is provided by the lattice constant and above integrals are to be understood as sums,
\beq \int^\Lambda \frac{d\qv}{(2\pi)^d} \rightarrow \frac{1}{V}\sum_{\qv,\qv\neq \bv{0}}\,.
\eeq
The sum runs over all permissible wavevectors on the lattice except the zero-mode, which must be excluded owing to the global conservation of the order parameter.
Note that $\Sigma^{(1)}$ is independent of the external wavevector $\kv$.

In a self-consistent treatment, the full correlation function $C$ is taken to be of the same form as $C_0$ but with renormalized parameters $r'$, $\kappa'$, that is
\beq C(\kv) = \frac{k_B T}{r'+\kappa' \kv^2 + O(k^2)}\,,
\eeq
Since $\Sigma$ itself depends on the renormalized $r'$ and $\kappa'$, eq.~\eqref{C_full} represents a system of two coupled integral equations for the determination of $r'$ and $\kappa'$ from the bare parameters $r$ and $\kappa$.
The wavevector-independent part of the self-energy, $\Sigma(\bv{0})$, obviously renormalizes the susceptibility parameter $r$,
\beq
r' = r+k_B T\Sigma(\bv{0})\,.
\label{r-renorm}
\eeq
The wavenumber-dependent part of $\Sigma$, which is of two-loop order, renormalizes the square-gradient parameter $\kappa$ and ultimately gives rise to a non-zero anomalous dimension $\eta$ at the critical point. At criticality, $\Sigma(\kv)$ scales as $k^{2-\eta}$.
Analogously, the fluctuation contributions to the coupling constant $u$ can be determined from the vertex-corrections to the four-point correlation function, which are also at least of two-loop order.
Taken together, one obtains a system of three coupled integral equations for $r'$, $\kappa'$ and $u'$ in dependence of the bare parameters.
For the present purposes, however, it is sufficient to focus only on the dominant effect, which resides in the renormalization of $r$. 
Using eqs.~\eqref{self1} and \eqref{self2}, the self-consistency equation for the renormalized temperature parameter $r'$ follows as
\begin{multline}
r' = r+ 3u k_B T \int^\Lambda \frac{d\qv}{(2\pi)^d} \frac{1}{r'+\kappa \qv^2}  -6u^2 (k_B T)^2 \\ \int^\Lambda \frac{d\qv_1}{(2\pi)^d}\frac{d\qv_2}{(2\pi)^d} \frac{1}{r'+\kappa \qv_1^2} \frac{1}{r'+\kappa \qv_2^2} \frac{1}{r'+\kappa (\qv_1+\qv_2)^2}\,,
\label{r-selfcon}
\end{multline}
which can easily be solved numerically. In practice, eq.~\eqref{r-selfcon} is used to find, for a given $r$ employed in a simulation, the corresponding value of $r'$, which will then allow one to compute the physical (renormalized) susceptibility $\chi = 1/r'$ and correlation length $\xi = (\kappa/r')^{1/2}$. This will give sufficiently accurate predictions in the crossover regime from mean-field to the critical region to be compared to simulation results.

\begin{figure}[t]
\centering
    (a)\includegraphics[width=0.65\linewidth]{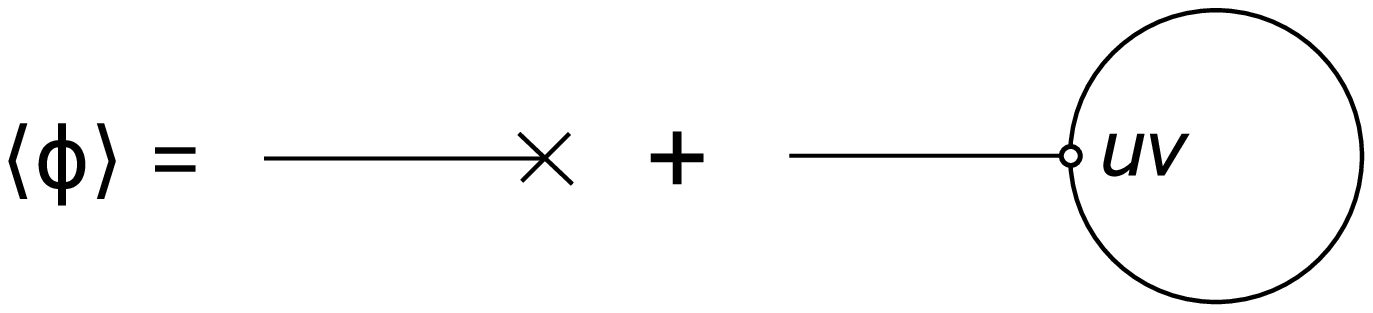}\\
    (b)\includegraphics[width=0.65\linewidth]{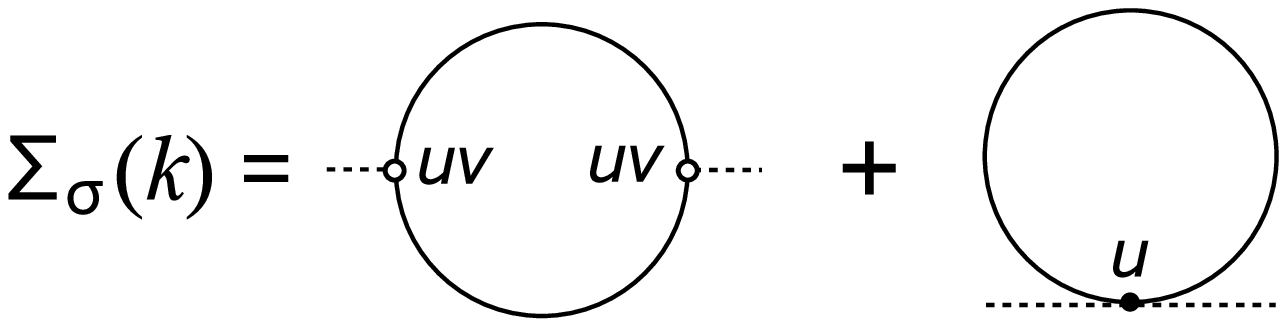}
   \caption{Perturbation theory in the broken phase: (a) Contributions to the expectation value of $\phi$ to $O(u)$. (b) Contributions to the self-energy of the order parameter in the broken phase, $\Sigma_\sigma$, to $O(u)$. In the broken phase, a new $\phi^3$ interaction with a coupling $uv$ appears. Note that, as $v\sim u^{-1/2}$, the first contribution to $\Sigma_\sigma$ is in fact of $O(u)$.}
    \label{fig:diagr-broken}
\end{figure}

Below the critical point, the order parameter acquires a non-zero expectation value $\bra \phi\ket$, which, at the mean-field level, is given by eq.~\eqref{phi0}. Fluctuation corrections, however, lead to a reduction of the mean-field expectation value. This effect can be isolated by splitting the order parameter as
\beq \phi(\rv) = v + \sigma(\rv)\,, \label{op-split}\eeq
where $v=\bra \phi\ket$ is enforced by requiring a vanishing expectation value of the fluctuation \cite{brezin_wallace_prl1972, brezin_wallace_prb1973}
\beq \bra \sigma \ket = 0\,. \label{vev-def}\eeq
Inserting eq.~\eqref{op-split} into the free energy functional \eqref{fef} leads to (up to an unimportant constant)
\begin{multline} \Fcal[v+\sigma] = \int d\rv\Big[ \frac{\kappa}{2}|\nabla\sigma|^2 + \frac{1}{2}(r+3 u v^2) \sigma^2 \\+ (rv+uv^3)\sigma + uv \sigma^3 + \frac{1}{4}u \sigma^4 \Big]\,.
\label{fef-broken}
\end{multline}
The last three terms can be considered as a perturbation around the Gaussian part given by the terms quadratic in $\sigma$ \cite{brezin_wallace_prl1972, brezin_wallace_prb1973}\footnote{To the order of perturbation expansion that will be considered here, a distinction between the bare and the renormalized $r$ in \eqref{fef-broken} is not necessary}.
To first non-trivial order, eq.~\eqref{vev-def} is represented in diagrammatic form by Fig.~\ref{fig:diagr-broken}a and follows as
\beq 0 = \bra \phi\ket = rv + uv^3 + 3uv \int \frac{d\qv}{(2\pi)^d} \frac{k_B T}{\kappa \qv^2+(r+3uv^2)}\,,
\label{vev-implicit}\eeq
which defines an implicit equation to be solved for the true $v$.
Note that the first two terms lead to the mean-field result for $v$, $v\st{MF}=(-r/u)^{1/2}$, while the last term gives the first-order fluctuation correction.
The correlation function of the shifted field $\sigma$ is obtained from eq.~\eqref{fef-broken} as
\beq C_\sigma(\kv) = \frac{k_B T}{\kappa\kv^2 + r_\sigma + k_B T\Sigma_\sigma(\kv)}\,,
\label{G-sigma}
\eeq
where $r_\sigma=r+3uv^2$ represents the inverse bare susceptibility of $\sigma$ and the leading-order self-energy corrections are given by the diagrams in Fig.~\ref{fig:diagr-broken}b, amounting to \cite{brezin_wallace_prl1972, brezin_wallace_prb1973, parisi_book, ma_book}
\begin{multline} \Sigma_\sigma(\kv) = -18 u^2 v^2 k_B T\int \frac{d\qv}{(2\pi)^d} \frac{1}{\kappa(\kv-\qv)^2+r_\sigma} \frac{1}{\kappa\qv^2+r_\sigma}\\ + 3 u \int \frac{d\qv}{(2\pi)^d} \frac{1}{\kappa\qv^2+r_\sigma}\,.
\label{sigma-broken}
\end{multline}
From eq.~\eqref{G-sigma} one obtains the true, renormalized susceptibility $\chi'=1/r'_\sigma$ with
\beq r_\sigma' = r_\sigma + k_B T \Sigma_\sigma(\bv{0}).
\label{rsigma-broken}
\eeq
Analogously to the situation in the symmetric state one could increase the accuracy of the perturbation expansion by replacing all appearances of $r_\sigma$ in the self-energy $\Sigma_\sigma(\bv{0})$ by $r_\sigma'$, thereby taking implicitly into account the fluctuation corrections to the correlation function given by the diagrams in Fig.~\ref{fig:diagr-broken}b to all orders.
However, to the order of perturbation theory set up in eq.~\eqref{sigma-broken}, the difference between the two expressions is negligible.

\subsection{Finite-size effects}\label{sec:fss}
On approaching the critical point in an infinite system, various intensive thermodynamic quantities display power-law divergences (see Table~\ref{tab:exp}). In a finite system, any quantity must necessarily stay finite and the critical divergences appear rounded \cite{privman_review1990, brankov_fss_book, amit_book}. Typically, deviations from the true critical behavior set in once the correlation length $\xi\propto \theta^{-\nu}$ of the hypothetical infinite system exceeds the system size $S$. In this case, one enters the so-called \emph{finite-size scaling} (FSS) regime, where the physical correlation length scales with the system size $S$.
Standard FSS theory \cite{privman_review1990, brankov_fss_book, amit_book}, which is summarized here, asserts that in this regime, the power-law dependence on the temperature of a thermodynamic observable $\Ocal$ in the infinite system, $\Ocal\sim \theta^{-x}$, essentially transfers to a power-law dependence on the system size
\beq \Ocal \sim S^{x/\nu} \tilde g_\Ocal(S/\xi) \sim S^{x/\nu} g_\Ocal(S^{1/\nu} \theta)\,,
\label{fss-ansatz0}
\eeq
where $g_\Ocal$ and $\tilde g_\Ocal$ are universal scaling functions \footnote{A universal scaling function can depend on its argument as $f(a_0 z)$ where $a_0$ is a non-universal constant \cite{amit_book}}.
To ensure that the correct asymptotic limit for the infinite system is reached, one must have $g_\Ocal(z)\sim A_{\Ocal \pm} |z|^{-x}$ as $z\rightarrow \pm\infty$ [where $A_{\Ocal \pm}$ denotes the corresponding amplitude, cf.\ eq.~\eqref{powerlaws}], while $g_\Ocal(z)$ must be regular for $z\rightarrow 0$. Note that, if $\Ocal$ represents the order parameter, only the limit $z\ra -\infty$ is relevant, since the order parameter is zero in the symmetric phase.
For the specific heat in 2D, FSS theory predicts that
\beq c_H\sim \log(S) g_C(S^{1/\nu}\theta)\,.
\eeq
In a finite system, the temperature value $r_{c,\Ocal}(S)$ for which an observable $\Ocal$ reaches a maximum (or, in case of the order parameter, vanishes) defines an \textit{apparent critical point}, which is typically found at a slightly different temperature $r$ than the critical point $r_c$ of the infinite system. The latter can be inferred by extrapolating the apparent critical point to the limit $S\ra\infty$ using $r_{c,\Ocal}(S)=r_c(1+ a_0 S^{-1/\nu})$, with a constant $a_0$ \cite{amit_book}. For the infinite system, all apparent critical points must merge.

In the present simulation approach, the density is globally conserved, implying that the global order parameter $m_S=\sum_i \phi_i/S^d$ and all derived quantities, such as the susceptibility,
$\chi_S = \frac{S^d}{k_B T} (\bra m_S^2\ket - \bra m_S\ket^2)\,$,
are trivially zero. Therefore, a standard FSS study based on the total system size $S$, as in eq.~\eqref{fss-ansatz0}, is not possible in this case.
Instead, ideas originally developed for grand-canonical simulations of the Ising-model \cite{binder_prl1981, binder_zphysb1981} -- which have been later successfully applied to canonical-ensemble simulations (employing Kawasaki-type dynamics) of lattice gas models and off-lattice fluids \cite{rovere_jpcm1990, rovere_zphysb1993} -- shall be followed here.
These methods essentially consist of dividing the total system of length $S$ into subsystems (blocks) of smaller length $L = S/2^i$ for integer $i$, and computing the quantities of interest in these subsystems\footnote{In principle, one could also allow arbitrary integer divisions of the system size instead of only powers of 2.}. In particular, a coarse-grained order parameter can be defined as
\beq m_L^{(b)} = \frac{1}{L^d}\sum_{i\in b} \phi_i\,,
\label{mL}
\eeq
where $i$ runs only over the lattice nodes that lie in the given subsystem $b$. For $L=1$, each block corresponds only to a single lattice site and thus $m_L^{(b)}$ becomes identical to the field variable $\phi_i$ at that site.
Note that, while the coarse-grained order parameter $m_L^{(b)}$ exhibits fluctuations, its average $\bra m_L\ket$ gives information only on the global asymmetry between the amounts of the two phases that are present; in particular, it can still be zero even in the phase-coexistence regime.
Thus, instead the quantity
\beq M_L \equiv \bra |m_L|\ket = \frac{1}{L^d}\big\bra \big| \sum_{i\in b} \phi_i \big|\big\ket
\label{ML} \eeq
is considered as the appropriate block order parameter for all temperatures, in agreement with the convention employed in the Monte-Carlo method \cite{binder_mc_intro}.
Note that, in a finite system, $M_L$ will be non-vanishing even in the disordered phase, but will approach zero in the thermodynamic limit $L\ra \infty$.
In eq.~\eqref{ML}, the average is performed over all blocks $b$ with the same size $L$ and over the statistical ensemble, or, alternatively, over time.

To define a susceptibility based on the coarse-grained order parameter, the disordered and ordered regimes have to be considered separately \cite{binder_mc_intro}. In the disordered phase, the standard definition
\beq \chi_L = \frac{L^d}{k_B T} (\bra m_L^2\ket - \bra m_L\ket^2) \qquad (\theta>0)\,
\label{chiL}
\eeq
is employed.
In the ordered phase, a slightly modified definition has to be used to ensure that $\chi_L$ only measures fluctuations around the equilibrium order-parameter value:
\beq \chi_L = \frac{L^d}{k_B T} (\bra m_L^2\ket - \bra| m_L|\ket^2)\qquad (\theta<0) \,.
\label{chiL-lowT}
\eeq
For deep quenches into the ordered regime, pronounced interfacial effects, however, prohibit a direct application of definition \eqref{chiL-lowT}. In these cases, we find here that the interfacial contributions have to be explicitly removed from the underlying order-parameter distribution (see below) in order to obtain a reliable estimate for the susceptibility.
It is important to realize that in the non-critical regime, where the correlation length is much smaller than $L$, above subbox susceptibility disagrees from the true susceptibility (as obtained, for instance, from the correlation function) by a boundary correction proportional $\sim \xi/L$ \cite{rovere_jpcm1990, rovere_zphysb1993}. 
Finally, in complete analogy to the susceptibility, a coarse-grained specific heat can be defined as
\beq
c_{H,L} = \frac{L^d}{k_B T} (\bra E_L^2\ket - \bra E_L \ket^2)\,,
\label{CL}
\eeq
where
$$E_L^{(b)}=\frac{1}{2L^d}\sum_{i\in b} \phi_i^2$$
is the average of the energy-like parameter field in a subbox.

When applied to the above block observables $\Ocal_L$, the original FSS ansatz, eq.~\eqref{fss-ansatz0}, must be extended by an additional scaling variable $L/S$ \cite{rovere_zphysb1993}. This is necessitated by the fact that, for $L=S$, order-parameter fluctuations are absent and, consequently, corrections to scaling are expected to depend on the ratio $L/S$. Thus, we can write
\beq \Ocal_L \sim L^{x/\nu} g_\Ocal(L^{1/\nu} \theta, L/S)\,,
\label{fss-ansatz}
\eeq
and similarly for the specific heat.
Strictly, this FSS ansatz is expected to be valid only for $\theta\rightarrow 0$ and $0\ll L \ll S$ with $L\rightarrow \infty$, while, outside this range, the influence of further corrections to scaling (for example, induced by the presence of irrelevant scaling fields) will become noticeable \cite{wegner_prb1972, privman_review1990}. However, in the case of Monte-Carlo simulations, it is often found that the simple FSS relation \eqref{fss-ansatz} works surprisingly well already for rather small lattice sizes \cite{binder_zphysb1981}.
The relation \eqref{fss-ansatz} will therefore be relied upon in this work as well.

The utility of the above FSS relations is based on the fact that they allow for a determination of universal critical parameters, such as exponents and amplitude ratios, provided that the location of the critical point is known. If the critical point is not known, one might still obtain reasonable estimates for the exponents and the critical temperature by trying different values until a good scaling of the data is achieved.
Relation eq.~\eqref{fss-ansatz} is particularly useful in a case where the exponents are already known and instead the critical point has to be located. In this case, one plots $\Ocal L^{-x/\nu}$ for different $L$ versus the coupling $u$ for a fixed value of $r$. By eq.~\eqref{fss-ansatz}, we have (neglecting the dependence on $L/S$)
\beq \Ocal L^{-x/\nu} \sim f_\Ocal(L^{1/\nu}\theta)\,,
\label{fss-crit}
\eeq
and thus, all curves cross through the same point when the critical coupling $u_c$ is passed \cite{goldenfeld_book, de_prd2005}.

\subsection{Order-parameter distribution}
Quantities such as the coarse-grained order parameter or susceptibility can be generally defined from the moments of an underlying order-parameter distribution function $P_L$ corresponding to a given subsystem size $L$ \cite{binder_prl1981, binder_zphysb1981},
\beq \bra m_L^k \ket = \int m^k P_L(m) dm\,. \eeq
The distribution function is particularly useful in the critical regime, where the order-parameter fluctuations have a pronounced non-Gaussian character that can not be fully captured by the low-order moments of $P_L$ alone.
In the phase-coexistence regime, the distribution moreover contains crucial information on surface tension \cite{binder_zphysb1981, binder_surf_pra1982} and phase-equilibria \cite{furukawa_binder_pra1982}.
Through its intimate connection to a coarse-grained free energy, it is also of fundamental importance in the description of nucleation and spinodal decomposition processes \cite{langer_metastable_1974, langer_baron_miller_pra1975, furukawa_binder_pra1982, binder_spinodals_pra1984}.
Practically, $P_L$ is obtained in a simulation by creating a histogram of the coarse-grained order parameter $m_L^{(b)}$ from all subsystems.
The above order-parameter distribution is therefore a coarse-grained quantity, and should thus not be confused with $P[\phi]$ of eq.~\eqref{prob-dist}, which is a functional of the order-parameter field $\phi_i$, that is $P[\phi] = P[\{\phi_i\}]$ depends on all values of $\phi$ on the lattice.
Formally, the coarse-grained distribution $P_L$ can be defined as a constrained average over all order-parameter fluctuations compatible with a given value of the average order parameter $m_L$ in a cell of size $L$,
\beq P_L(m) = \frac{1}{Z}\int \prod_i d\phi_i \,\delta \left(m - \frac{1}{L^d}\sum_{i\in b} \phi_i\right) e^{-\Fcal[\phi]/k_B T}
\label{PL-def}
\eeq
where $Z$ is a normalization factor.
For instance, for $L=1$, the distribution $P_L$ is obtained by integrating the probability functional $P[\{\phi_i\}]$ over all $\phi_i$ except one,
$P_1(\phi_j) = \int \prod_{i\neq j} d\phi_i P[\{\phi_i\}]$.

The connection of the constrained distribution $P_L(m)$ of eq.~\eqref{PL-def} to thermodynamics can be made explicit by defining a \textit{constrained Helmholtz free energy} $F_L(m)$ via \cite{binder_zphysb1981, newlove_bruce_jphysa1985, radescu_1995, troester_prb2005}
\beq F_L(m) = -k_B T\log (P_L(m) Z) = -k_B T \log P_L(m) + F\,,
\label{FL-def}
\eeq
where relation \eqref{HelmholtzF} has been used.
The thermodynamic Helmholtz free energy $F$, eq.~\eqref{HelmholtzF}, is obtained from $F_L$ by
\beq \exp(-F/k_B T) = \int_{-\infty}^{+\infty} dm \exp[-F_L(m)/k_B T]\,.
\eeq
It must be emphasized that $F_L$ is in general different from the bare Landau potential $f_0$ [eq.~\eqref{landau-pot}], as the former is derived from an integral over the full probability functional and therefore includes (due to the gradient term) also the effects of interactions between the fluctuations.
Exceptions are the limit $T\ra 0$ as well as $F_1$ in the Ising limit, since in these cases the interaction between different cells can be neglected.
In the large volume limit, the constrained free energy can be shown to be equivalent to the so-called effective potential often employed in field theory \cite{oraifeartaigh_effpot_npb1986}.
Above the critical point, the constrained free energy is a direct measure for fluctuations of the coarse-grained order parameter $m_L$ around the equilibrium state. Below the critical point this picture breaks down, as $P_L$ will then not only receive contributions from homogeneous fluctuations but also from the presence of two-phase states. Nevertheless, $F_L$ is often found to be well approximated by a simple Landau form, i.e.\ a low-order polynomial in $m$ \cite{kaski_binder_prb1984, tsypin_prl1994, radescu_1995, tsypin_prb1997, troester_prb2005}.
An often invoked alternative characterization of the order-parameter distribution that avoids fitting a potential is based on its higher-order cumulants \cite{binder_zphysb1981}. However, it is found that, in the present case, the standard cumulant ratio $U_3$ does not appear to have a well-defined $L$-independent limit at the critical point. This is most likely caused by interfacial effects, as argued in \cite{rovere_zphysb1993}.
The cumulant analysis will therefore not be pursued in the present work.

The shape of the coarse-grained distributions $P_L$ and the corresponding free energies $F_L$ can be anticipated based on simple physical arguments \cite{binder_zphysb1981, kaski_binder_prb1984}.
Far above the critical point, non-linear effects are small and thus the order-parameter distribution is expected to be well approximated by a Gaussian centered around the average order-parameter value $\bra m \ket = 0$ \cite{binder_zphysb1981},
\beq\begin{split}
P_L(m) &= \frac{1}{(2\pi\bra m_L^2 \ket)^{1/2}} \exp\left(-\frac{m^2}{2\bra m_L^2\ket}\right)\\
&= \frac{L^{d/2}}{(2\pi k_B T \chi_L)^{1/2}} \exp\left(-\frac{m^2 L^d}{2k_B T \chi_L}\right)\,,
\label{PL-highT}
\end{split}\eeq
where relation \eqref{chiL} for the variance $\bra m_L\ket^2$ has been used.
The width of each Gaussian decreases with larger coarse-graining length $L$ as more and more fluctuations are averaged out and the distribution approaches the high-temperature fixed-point.

Below the critical point, the shape of $P_L$ depends distinctly on the size of the coarse-graining length $L$ in comparison to the correlation length $\xi$.
For the case of a phase transition at the critical density (which is exclusively considered here), the global conservation of the order parameter requires that, below the critical point, equal volumes of liquid and vapor are present in the simulation box.
For $L\lesssim \xi$, a given subbox will typically cover either liquid or vapor and thus $P_L$ is expected to show two approximately Gaussian peaks centered around the spontaneous values of the order parameter $\pm M_L$ [which, to a first approximation are given by eq.~\eqref{phi0}],
\begin{multline}
P_L(m) = \frac{1}{2}\frac{L^{d/2}}{(2\pi k_B T \chi_L)^{1/2}} \Big[ \exp\left(-\frac{(m-M_L)^2 L^d}{2k_B T \chi_L}\right) \\+ \exp\left(-\frac{(m+M_L)^2 L^d}{2k_B T \chi_L}\right)\Big]\,.
\label{PL-lowT}
\end{multline}
In general, the region between the peaks of the distribution, $-|M_L| <m < |M_L|$, represents the probability not only for homogeneous order-parameter fluctuations, but also for the occurrence of two-phase configurations in a subsystem.
For $L\lesssim \xi$, the box cannot cover complete phase-separated states, and thus for this case the height of $P_L(0)$ is essentially a measure for the probability of homogeneous fluctuations.
In general, however, homogeneous fluctuations are exponentially suppressed by the volume $L^d$ [eq.~\eqref{PL-lowT}], in contrast to heterogenous fluctuations, whose free energy cost is just proportional to the area of the interface, $L^{d-1}$.
Thus, for subsystems with $L\gg\xi$, homogeneous fluctuations give a completely negligible contribution to the central region of $P_L$, and one can estimate the probability for a heterophase fluctuation as
\beq P_L(0)\simeq \text{const} L^x\exp \left(-\frac{2L^{d-1}\sigma}{k_B T}\right)\,,
\label{PL0-surf}
\eeq
with an empirical exponent $x$ that is typically found to be close to zero \cite{binder_surf_pra1982, berg_interf_prb1993}.
In the ensemble average, liquid and vapor phases will occur equally often and with any proportion in each subbox. Thus, one expects that for large $L$ the central region of $P_L$ will become flat and, in the limit $L\rightarrow \infty$, where interfacial contributions become negligible, eventually attain the same level as the peaks. This is also expected based on the notion of a coarse-grained free energy $F_L$, as the true free energy in the thermodynamic limit, $F_\infty$, must be convex due to reasons of stability (as also implied by the Maxwell construction) \cite{binder_firstorder_review1987, phase_trans_mater_book}.

In the vicinity of the critical point, that is, in the FSS region characterized by $\xi\gtrsim L$, the distribution function becomes markedly non-Gaussian and one can make a scaling ansatz,
\beq P_L(m) = L^{\beta/\nu} \tilde P(m L^{\beta/\nu}, L^{1/\nu} \theta, L/S)\,,
\label{PL-fss}
\eeq
with $\tilde P$ being a universal scaling function \cite{binder_prl1981, binder_zphysb1981, rovere_zphysb1993, amit_book}. As can be easily checked, the above relation reduces to the corresponding FSS relations for the moments, eq.~\eqref{fss-ansatz}, if, additionally, use of the hyperscaling relation $d\nu=\gamma+2\beta$ is made.
It must be emphasized that the scaling function $\tilde P$ is universal only for sufficiently large $L$, where it embodies the collective features of the critical phase transition. For small $L$, in fact different shapes for $P_L$ at criticality are possible, depending on the location on the critical line: Toward the displacive limit, the barrier between the minima of the local potential is low, leading to a nearly Gaussian shape of $P_L$, in contrast to a pronounced double-peak structure in the Ising limit, where $P_L$ closely reflects the on-site potential $f_0$ \cite{bruce_jphysc1981}.
The large-scale properties of the order-parameter distribution (in 2D and 3D) have been extensively studied in previous works via field theoretic approaches \cite{bruce_jphysc1981, newlove_bruce_jphysa1985, eisenriegler_tomaschitz_prb1987, esser_physa1995, chen_physA1996} and Monte-Carlo simulations of Ising-like systems \cite{binder_zphysb1981, kaski_binder_prb1984, tsypin_prl1994, hilfer_wilding_jphysa1995, radescu_1995, troester_prb2005}.


\section{Simulation method}
\label{sec:method}
\subsection{Model}
In the present work, the equilibrium behavior of the Ginzburg-Landau model is simulated on a square lattice via a fluctuating hydrodynamics approach. Specifically, the Langevin extension of the non-ideal fluid LB model of Swift et al.\ \cite{swift_prl1995, swift_pre1996}, introduced in \cite{gross_flb_2010}, is employed.
The LB equation (LBE) can be understood as a discretization of the continuum Boltzmann equation and contains as a subset the Navier-Stokes equations in the limit of long time and length scales \cite{succi_book}.
The LBE describes the evolution of a set of distribution functions $f_i(\rv) \equiv f(\rv,\cv_i)$ on a lattice streaming along a finite number of possible velocity directions $\cv_i$ linking the nodes. Here, simulations are performed on a D2Q9 lattice, that is, the space dimension is $d=2$ and $i=1,\ldots,9$.
Employing a simple BGK-approximation to the collision operator, the present LB model is defined by the evolution equation
\begin{multline}
 f_i(\rv +\bv{c}_i\Delta t, t+\Delta t) \\= f_i(\rv,t) - \frac{\Delta t}{\tau}[f_i(\rv,t) - f_i\ueq(\rv,t)] + \vartheta_i(\rv,t)\,,
\label{lbe}
\end{multline}
where $t$ is the time, $\Delta t$ is the time step, $\tau$ is a relaxation time, $f_i\ueq$ is the equilibrium distribution and $\vartheta_i$ is a random force term, to be specified below.
The relevant observable quantities are given by the low-order moments of the distribution function. In particular, we have for the density $\rho$ and the fluid velocity $\uv$,
\beq \rho = \sum_i f_i = \sum f_i\ueq,\quad \rho\uv = \sum_i f_i \cv_i = \sum_i f_i\ueq \cv_i\,.
\label{mom}
\eeq
In the model of Swift et al.\ \cite{swift_prl1995, swift_pre1996}, equilibrium thermodynamics as embodied by the Ginzburg-Landau free energy functional is implemented by requiring the second moment of the equilibrium distribution to recover a thermodynamic pressure tensor $\Pt$,
\begin{multline} \sum_i c_{i\alpha} c_{i\beta} f_i\ueq = P_{\alpha\beta} + \rho u_\alpha u_\beta \\
+ \nu(u_\alpha\partial_\beta\rho + u_\beta \partial_\alpha\rho + u_\gamma \partial_\gamma\rho\delta_{\alpha\beta})\,.
\label{holdych-cond}
\end{multline}
The term proportional to the kinematic viscosity $\nu \equiv \eta/\rho = (\tau-1/2)/3$ is introduced in the above equation to improve Galilean invariance \cite{holdych_gal_1998}.
The pressure tensor $\Pt$ is given by
\beq P_{\alpha\beta} = \left(p_0 - \kappa \rho \nabla^2 \rho - \frac{\kappa}{2}|\nabla\rho|^2\right) \delta_{\alpha\beta} + \kappa (\nabla_\alpha \rho)(\nabla_\beta \rho)\,,
\label{press-th}
\eeq
where $p_0=\rho \frac{1}{\rho_0}\partial_\phi f_0-f_0$ is the thermodynamic pressure.
The pressure tensor satisfies the relation $\nabla\cdot\Pt = \rho\nabla (\delta \Fcal /\delta \phi)$
and can be obtained from the free-energy functional \eqref{fef}, for instance, via the Noether theorem, the principle of least action or from the requirement of hydrostatic equilibrium \cite{anderson_diffuse_1998, briant_pre2004, jasnow_vinals_1996, yang_molecular_1976}.
Physically, it accounts for the energetic balance between changes in fluid structure due to advection and surface-tension \cite{jacqmin_jcomp1999} and thus ensures that the equilibrium order-parameter distribution, eq.~\eqref{prob-dist}, remains unchanged by the flow \cite{onuki_book}.
The explicit expression for the modified-equilibrium distribution $f\ueq_i$ on a D2Q9 lattice used in the present work is given by \cite{pooley_spurious_2008}
\begin{multline} f\ueq_i = w_i\Big[\rho u_\alpha c_{i\alpha} + \frac{3}{2}\left( c_{i\alpha} c_{i\beta} - \frac{1}{3} \delta_{\alpha\beta}\right) \big[\rho u_\alpha u_\beta + \\
\nu(u_\alpha \pd_\beta \rho + u_\beta \pd_\alpha \rho u_\gamma \pd_\gamma\rho \delta_{\alpha\beta} )\big] + p_0 - \rho\nabla^2\rho \Big] + \\
w_i^{xx} \kappa (\pd_x\rho \pd_y\rho) + w_i^{yy}\kappa (\pd_y\rho)(\pd_y\rho) + w_i^{xy}\kappa (\pd_x\rho)(\pd_y\rho)\,,
\end{multline}
with the weights taken as $w_{1-4}=1/3$, $w_{5-8}=1/12$, $w_{5-8}^{xx}=w_{5-8}^{yy}=-1/24$, $w_{1,2}^{xx}=w_{3,4}^{yy}=1/3$, $w_{3,4}^{xx}=w_{1,2}^{yy}=-1/6$, $w_{1-4}^{xy}=0$, $w_{5,6}^{xy}=1/4$ and $w_{7,8}^{xy}=-1/4$.

To complete the description of the employed LB model, the properties of the noise variables $\vartheta_i$ have to be specified.
In order to properly account for mass and momentum conservation, the moment representation of the LBE is invoked \cite{dHumieres_MRT_1992, benzi_physrep1992}. This representation is defined by a set of basis vectors $T_{ai}$ ($a=1,\ldots,9$) that admit the distribution function $f_i$ to be expanded in terms of a set of moments $m_a$ as
\beq f_i(\rv,t) = T_{ai} \frac{w_i}{N_a} m_a(\rv,t)\,,
\eeq
where the $N_a$ are the squared lengths of the basis vectors $\bv{T}_{a}$. In the present work, the $\bv T_{a}$ as given in \cite{gross_flb_2010} are used. The first three moments $m_{a=1,2,3}$ then follow as $\rho,\rho u_x$, and $\rho u_y$, while the higher moments cover the stresses ($a=4,5,6$) and the so-called ghost-modes ($a=7,8,9$).
The moment representation of the noise is equivalently defined as $\hat\vartheta_a \equiv T_{ai} \vartheta_i$. Only the expression for $\hat \vartheta_a$ is stated below, which has a much simpler form than $\vartheta_i$.

As shown in \cite{gross_flb_2010}, due to the use of a modified-equilibrium distribution to incorporate the non-ideal gas thermodynamics [see eq.~\eqref{holdych-cond}], the noise obtained from the fluctuation-dissipation theorem of the LBE is wavenumber-dependent. This is clearly undesirable, as such a form of noise is not directly applicable to spatially inhomogeneous situations involving phase-separation. However, as the offending terms in the noise covariance are proportional to the square-gradient parameter $\kappa$, it is possible, by reducing $\kappa$ appropriately, to employ spatially uncorrelated noise while still maintaining satisfactory equilibration.
Additionally, spatially uncorrelated noise has the advantage of being straightforward to implement and computationally cheap.

The noise is thus taken to be a Gaussian random variable without explicit correlations in space or time
\beq
\bra \hat \vartheta_a(\rv,t) \hat \vartheta_b(\rv',t')\ket = \Theta_{ab}(\rv) \delta_{\rv,\rv'}\delta_{t,t'}\,.
\eeq
However, in order to properly account for spatially inhomogeneous fluid properties -- occurring, for instance, in the phase-coexistence regime -- the covariance $\ten \Theta$ shall be allowed to depend locally on position \cite{gross_dbe_2011}. The final expression for the noise covariance, taking into account above-mentioned modification and neglecting terms proportional to the square-gradient parameter $\kappa$, is obtained as \cite{gross_flb_2010}
\begin{widetext}
\beq
\bv{\Theta}(\rv) = \frac{3\rho(\rv) k_B T}{\Delta V\Delta t} \frac{1}{\tau}\left(2-\frac{1}{\tau}\right) \left(
\begin{array}{ccc|ccc|ccc}
 . & . & . & . & . & . & . & . & . \\
 . & . & . & . & . & . & . & . & . \\
 . & . & . & . & . & . & . & . & . \\
\hline
 . & . & . & 4 \left[2-3c_s^2(\rv)\right]  & . & . & . & . & 12 \left[c_s^2(\rv)-\frac{1}{3}\right]  \\
 . & . & . & . & 4/9  & . & . & . & . \\
 . & . & . & . & . & 1/9 & . & . & . \\
\hline
 . & . & . & . & . & . & 2/3 & . & . \\
 . & . & . & . & . & . & . & 2/3 & . \\
 . & . & . & 12 \left[c_s^2(\rv)-\frac{1}{3}\right] & . & . & . & . & 16\left[\frac{5}{4} - \frac{3}{4} c_s^2(\rv)\right] \\
\end{array}
\right).
\label{lbnoise}
\eeq
\end{widetext}
Here, $\Delta V$ is the volume of a lattice cell, which, as well as the time step $\Delta t$, is equal to one in lattice units (l.u.).
The quantity $k_B T$ fixes the fluctuation amplitude and is essentially a free parameter of the model, subject only to the low-Mach number constraint of the LB method.
The above noise covariance ensures that that fluctuations of the fluid velocity obeys locally the equipartition theorem of statistical mechanics,
\beq \bra u_\alpha(\rv) u_\beta(\rv')\ket = \frac{k_B T}{\rho(\rv)} \delta_{\alpha\beta}\delta_{\rv,\rv'}\,.
\label{vel-correl}
\eeq
In a simulation, Gaussian noise with a non-diagonal covariance matrix can be created via a Cholesky-transform (see \cite{gross_flb_2010}).

As a consequence of the LB dynamics of eq.~\eqref{lbe}, at large length and time scales the density and momentum obey a continuity equation,
\beq \partial_t \rho + \nabla\cdot (\rho \uv) = 0\,.
\label{contin-eq}
\eeq
and a momentum-conservation (Navier-Stokes) equation for a non-ideal fluid,
\beq
 \partial_t(\rho \uv) + \nabla \cdot (\rho\uv\uv) =  -\nabla \cdot \Pt + \nabla\cdot \ten \sigma + \nabla\cdot\ten R\,,
\label{nse}
\eeq
where
\beq \sigma_{\alpha\beta} = \eta\left(\pd_\alpha u_\beta + \pd_\beta u_\alpha - \frac{2}{d} \pd_\gamma u_\gamma\right) + \zeta \pd_\gamma u_\gamma \eeq
is the viscous stress tensor,
\beq \eta=\frac{\rho}{3}\left(\tau-\onehalf\right)
\eeq
is the shear viscosity (which should not be confused with the anomalous dimension index) and
\beq \zeta=\frac{\rho}{3}\left(\tau-\onehalf\right)\left(2-3 c_s^2\right)
\label{bulkvisc}
\eeq
is the bulk (or volume) viscosity.
It is noteworthy that, in the BGK-approximation to the Boltzmann equation, the (bare) viscosities depend on the local density \cite{dellar_visc_2001}.
In the critical regime, however, it always possible to choose the parameters in the Landau free energy such that the magnitude of the density fluctuations is small compared to the background density, $\delta\rho/\rho\ll 1$ (cf.~Fig.~\ref{fig:ophist-crit}). If the viscosities are approximated as constants, and furthermore the non-linear advection term (which is not relevant at criticality \cite{hohenberg_halperin}) is neglected, the Navier-Stokes equation simplifies to
\begin{multline}
 \partial_t(\rho \uv) = -\nabla \cdot \Pt + \eta \nabla^2\uv + \left(\zeta +[1-2/d]\eta \right)\nabla\nabla\cdot\uv +\nabla\cdot \ten{R}\,.
\label{nse2}
\end{multline}
Note that expression \eqref{bulkvisc} for the bulk viscosity differs from the standard LB expression by a factor of $(2-3 c_s^2)$, which is an artifact of the modified equilibrium distribution of the present LB model \cite{swift_pre1996, gross_flb_2010} \footnote{Terms proportional to $\kappa$ have been neglected in eq.~\eqref{bulkvisc}}.

The random stress tensor $\ten{R}$ imparts thermal noise on the fluid momentum which is then transferred to the order-parameter sector, leading -- in equilibrium -- to thermal fluctuations of $\phi$ according to the distribution \eqref{prob-dist}.
The random stress tensor, which is directly related to the LB noise variables $\vartheta_i$, is a Gaussian white noise source with correlations given by
\begin{multline}
 \bra R_{\alpha \beta}(\bv{r},t) R_{\gamma \delta}(\bv{r'},t') \ket = 2k_B T \Big[ \eta(\rv) \Big(\delta_{\alpha \gamma} \delta_{\beta \delta} + \delta_{\alpha \delta} \delta_{\beta \gamma} \\- \frac{2}{d} \delta_{\alpha \beta}\delta_{\gamma \delta}\Big)
+ \zeta(\rv) \, \delta_{\alpha \beta}\delta_{\gamma \delta}\Big]\delta(\bv{r}-\bv{r'})\delta(t-t')
\,.
\label{rand-stress-correl}
\end{multline}
The above expression is identical to the standard Landau-Lifshitz result for ideal fluids \cite{landau_fluidmech}, except for the locally varying viscosities that are needed to properly take into account possible spatial inhomogeneities of the fluid.
Note that additional error terms in the Navier-Stokes eq.~\eqref{nse} originating from the LB model have been neglected. These terms generally depend by a positive power on the density gradient or flow velocity \cite{holdych_gal_1998} and are expected to be negligible in the  present case.

\subsection{Setup}
Simulations in the critical regime require fine-tuning of parameters as implied by the phase-diagram (Fig~\ref{fig:phase-diag}) as well as by LB-specific constraints.
First of all, since the flow velocity must not exceed the lattice sound speed $\cslb$ (where $\cslb^2=1/3$ for D2Q9 models), expression \eqref{vel-correl} for the fluctuation variance, $k_B T/\rho_0 = \bra u_\alpha^2 \ket$, directly implies
\beq k_B T \ll \cslb^2 \rho_0 \,. \label{eq:low-mach}\eeq
Second, the density must remain strictly positive. Due to the Gaussian character of the density fluctuations this implies that the average density fluctuation should remain much smaller than the mean density. Neglecting, for simplicity, spatial correlations, density fluctuations have a variance of $\bra \Delta\rho^2\ket =\rho_0 k_B T/c_s^2$; hence, requiring that $\bra \Delta \rho^2\ket^{1/2}\ll \rho_0$ leads to
\beq k_B T \ll c_s^2\rho_0\,, \label{den_fluct_constr}\eeq
which is a more stringent constraint than eq.~\eqref{eq:low-mach}, since $c_s^2\sim \chi^{-1} \ll \cslb^2$ for a non-ideal fluid in the critical regime.
Thus, the fluctuation temperature $T$ must be chosen sufficiently small for the velocity fluctuations not to violate the approximate incompressibility of the LB method. Typically, values of $k_B T=10^{-7}$ l.u.\ or less are sufficient.

In the critical regime, it is particularly important to ensure accurate equilibration of the fluid at all scales, since here mode-coupling effects are dominant and thus errors induced at the smallest scales can propagate to larger ones and possibly infect the whole simulation.
Since the noise covariance \eqref{lbnoise} was derived in the limit of $\kappa\ra 0$, it is expected that using a sufficiently small value for $\kappa$ ensures equilibration to high accuracy. In fact, it is found that values of $\kappa\lesssim 10^{-3}$ are already sufficient in the present case.
This is demonstrated in Fig.~\ref{fig:momtm-equil}, where -- for a set of parameters in the critical region (see below) -- the Fourier-transform of the spatial correlation function of the fluid momentum $\bv{j} = \rho \uv$ is compared to the theoretically expected result [eq.~\eqref{vel-correl}], $\bra |j_\alpha(\kv)|^2 \ket = \rho_0 k_B T$. To achieve a reasonable statistical accuracy, correlations have been computed by averaging over $5000$ snapshots over a total simulation time of $10^7$ timesteps. As the figure shows, perfect equilibration at all scales with an error below a few percent is obtained.
\begin{figure}[t]
\centering
    \includegraphics[width=0.75\linewidth]{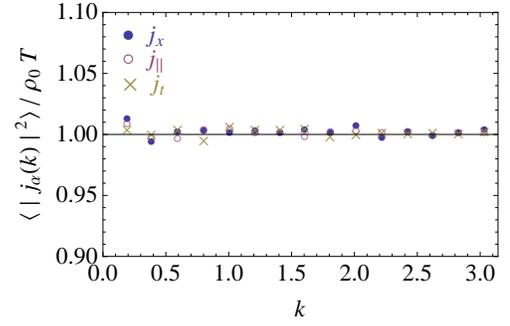}
   \caption{(Color online) Equilibration of momentum for a critical fluid. $j_x$ denotes the $x$-component of the momentum, while $j_{||}$ and $j_\perp$ denote the projections of $\bv{j}$ longitudinal and transversal to the wavevector $\kv$. Simulation parameters (in l.u.): $\kappa=10^{-4}$, $r=4.96\times 10^{-5}$, $u=5.16\times 10^{-2}$, $k_B T=10^{-7}$. All LB relaxation times are set to $\tau=0.8$.}
    \label{fig:momtm-equil}
\end{figure}

The chosen value of $\kappa$ leads to restrictions on the possible values of the free energy parameters $r$ and $u$ in the critical region, as these quantities enter the reduced $\tilde r$ and $\tilde u$ [eq.~\eqref{reduced-param}] which span the phase-diagram (Fig.~\ref{fig:phase-diag}) of the Ginzburg-Landau model. In order to observe Ising-type critical behavior, one would want to avoid too close proximity to the displacive limit and hence choose a large value for $\tilde r$ \cite{milchev_1986, amit_book}.
As outlined in \cite{wagner_interface_2007}, for the deterministic LB method there exists, as a consequence of the finite-difference approximations to the spatial derivatives, a lower limit for the interface width of around 1~l.u., below which LB simulations produce potentially wrong dynamics even for small density differences.
Since the large-scale fluctuations at the critical point essentially consist of phase-separated regions (``critical droplets'') that continuously break-apart and reconnect \cite{fisher_dropletmodel}, it is plausible that the restriction on the interface width derived for the non-fluctuating case continues to hold in some form also for fluctuating critical domains.
Since the interface width of the critical domains is roughly given by the mean-field result
\beq w\simeq \sqrt{-2\kappa/r} = \sqrt{2/\tilde r}\,,\eeq
one obtains an upper bound on $\tilde r$ of around 2 l.u.
Crucially, this argument implies that one can not get arbitrarily close to the Ising limit without sacrifying correct dynamics.
Simulation results obtained in this work, however, indicate that this appears to be not too severe of a restriction for the application of the LB method to critical phenomena.
Taking, for instance, $k_B T=10^{-7}$, $w\sim O(1)$ and $\kappa\sim O(10^{-4})$, a simulation in the critical regime then requires that $\tilde r_c\sim O(1)$ and $\tilde u_c\sim O(1)$, implying $r_c\sim O(10^{-4})$ and $u_c\sim O(10^{-1})$.
To obtain a more precise location of the critical point, one can gradually change one of the parameters $r$, $u$, or $T$ and visually inspect the density field, compute the structure factor or apply a FSS analysis, as shown below.
Note that, in principle, one can traverse the critical regime by either changing $T$, $u$ or $r$, keeping in each case the other parameters fixed. To stay in line with the usual field-theoretical notion of the temperature-like variable, usually only $r$ will be varied in the present work.

In the critical region, relaxation processes become extremely slow (\textit{critical slowing down}), requiring, especially at long wavelengths, a large simulation time in order to collect a sufficiently large number of statistically independent samples (cf.~\cite{binder_mc_intro}).
Specifically, in an isothermal critical fluid, density fluctuations relax via overdamped sound waves (to be discussed in more detail in a forthcoming publication \cite{gross_critical_dynamics}) which decay with a rate of $\Gamma(k)=c_s^2(k)/\nu_l$, where the generalized speed of sound is given by (see, e.g., \cite{gross_flb_2010}) $c_s^2(k)=c_s^2+\rho\kappa k^2$ and $\nu_l=(\eta+\zeta)/\rho$ is the longitudinal viscosity for a 2D fluid.
Consequently, the largest possible relaxation time of the order parameter can be estimated as
$t_\rho \sim \nu_l/c_s^2(k\st{min}) \sim S^2 \nu_l/(c_{s,0}^2+ 4\pi^2\rho \kappa )$, with $S$ being the system size, $k\st{min}=2\pi/S$ the minimum wavenumber and $c_{s,0}^2$ the value of $c_s^2$ for a correlation length of $\xi=1$. Here, the critical (mean-field) FSS of the thermodynamic speed of sound, $c_s^2\propto 1/\chi = c_{s,0}^2 \xi^{-2} \propto S^{-2}$, has been used.
Thus, the performance of a simulation can be optimized by choosing a small value of $\nu_l$. (Note, however, that the momentum relaxation time scales as $\propto S^2/\eta$.)
For instance, $S=256$, $\kappa=10^{-3}$ and $\nu_l=10^{-2}$ gives an order-parameter relaxation time of $t_\rho\sim 10^{6}$ l.u.
Hence, a simulation must run roughly $10^8$ timesteps until accurate statistical information (errors less than 1\%) for order-parameter related quantities is obtained when averaging over a few hundred realizations. This requirement of simulation time might seem excessive, but one should keep in mind that,
due to the global conservation of the order parameter, large-wavelength fluctuations require the rearrangement of mass over large distances.


\section{Results}
\subsection{Correlation function}

\begin{figure*}[t]
\centering
    (a)\includegraphics[width=0.4\linewidth]{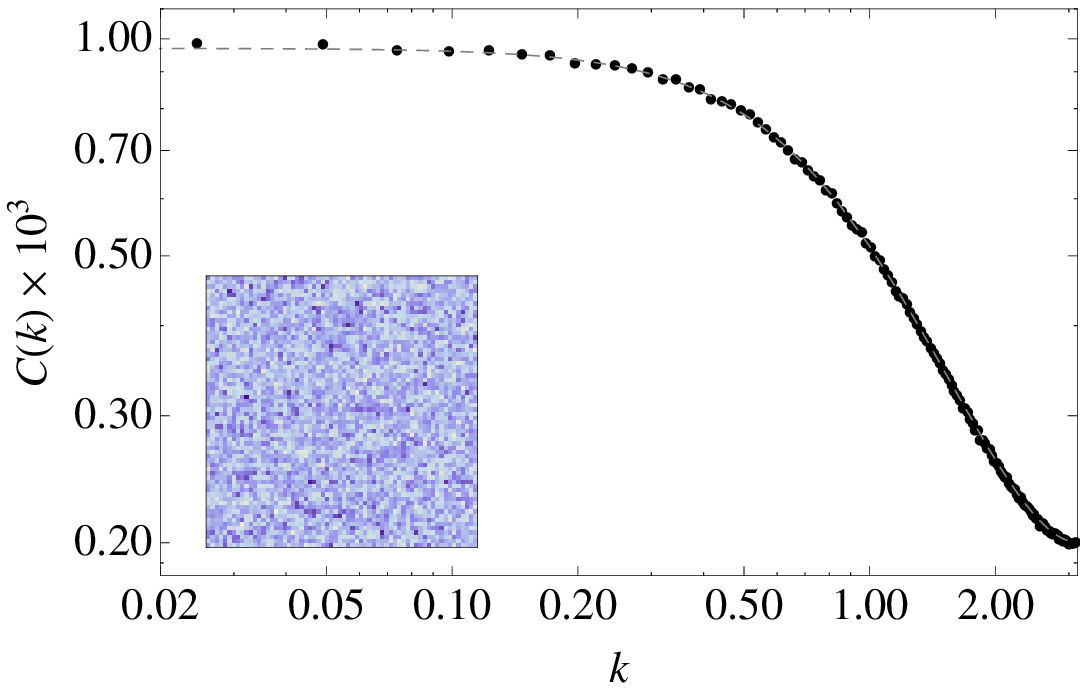}
    (b)\includegraphics[width=0.41\linewidth]{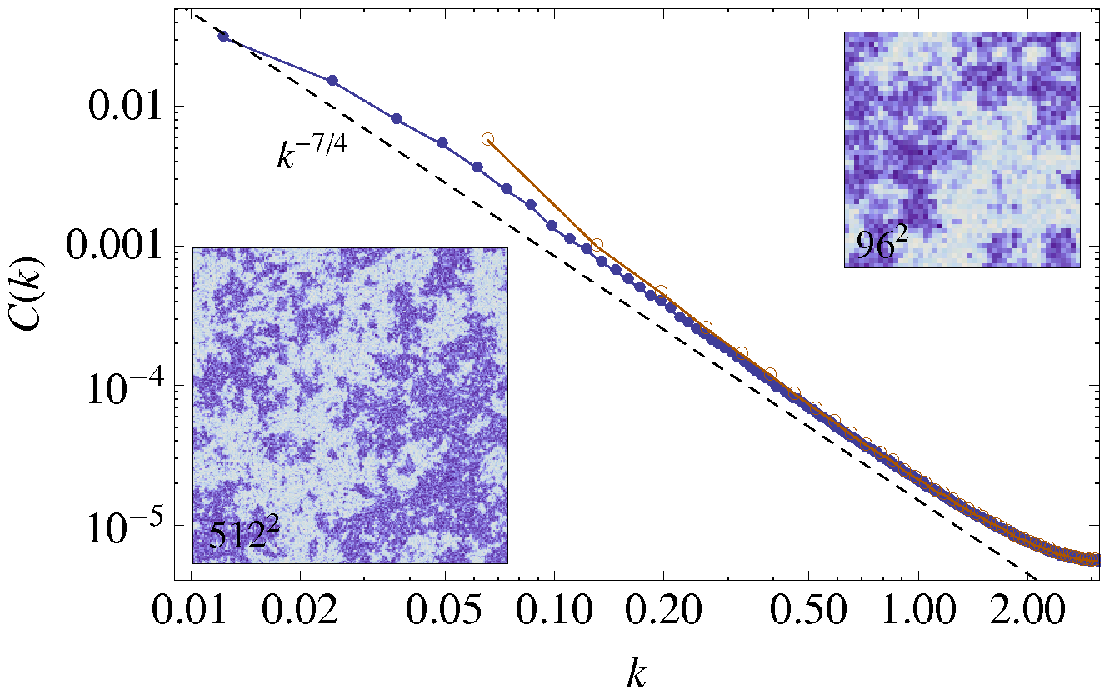}
    \caption{(Color online) Correlation function $C(k)$ vs.\ wavenumber $k$ obtained from a simulation (a) far above and (b) close to the critical point. The insets show the representative order-parameter field. In (a), the dashed curve represents a fit to an Ornstein-Zernike form [eq.~\eqref{CMF}], while in (b), it represents the critical power law $k^{-2+\eta}$. In (b), the structure factor and order-parameter field for two different system sizes, $S$=256 ($\bullet$) and 96 ({$\circ$}). Note the finite-size effects at low $k$. Simulation parameters: (a) $r=10^{-4}$, $u=4\times 10^{-3}$, $\kappa=10^{-4}$, $k_B T=10^{-7}$; (b) $r=-10^{-4}$, $u=4$, $\kappa=2.0\times 10^{-5}$, $k_B T=5.31\times 10^{-10}$ (all in l.u.). }
    \label{fig:sfac}
\end{figure*}
In the simulations, the structure factor is computed from the order-parameter field $\phi$ on the discrete lattice by
\beq C(\kv) = \frac{1}{N}\left\bra \left|\sum_\rv \phi(\rv) e^{-\im \kv\cdot \rv}\right|^2\right\ket\,,
\label{G-comput}
\eeq
where the brackets indicate time-average over many statistically independent samples and $N$ is the total number of lattice points.
Due to periodic boundary conditions and the real-valuedness of $\phi$, it suffices to consider the structure factor in the first half of the first Brillouin zone, that is, in the wavenumber range $0<k<\pi$. Global mass conservation enforces $C(\bv{0})=0$ (this point is excluded from the plots). Figure~\ref{fig:sfac} shows the structure factor together with sample snapshots of the order-parameter field above and at the critical point. Error bars are of the order of the symbol size and not shown.

Above the critical point (Fig.~\ref{fig:sfac}a), the correlation function assumes a simple Ornstein-Zernike form, eq.~\eqref{CMF}.
In eq.~\eqref{CMF}, $k^2$ should be understood as the Fourier-transformed \emph{discrete} Laplacian, which reveals itself in a deviation of the high-$k$-part of $C(k)$ from a simple $k^{-2}$ power-law expected in the continuum case (see, e.g., \cite{gross_flb_2010}). The discrete lattice effect becomes noticeable for wavenumbers $k\gtrsim 1$.
Note that around $r=0$, self-energy corrections suppress the correlation length and compressibility below their mean-field values given by eqs.~\eqref{correlMF}, \eqref{chiMF}.

At the critical point, the correlation function is expected to assume a power-law, $C(k)\sim k^{-2+\eta}$ [eq.~\eqref{C-crit}], for $k\gtrsim 1/\xi$.
In the above equation, $\eta$ is the anomalous-dimension exponent, which takes a value of $\eta=1/4$ for the 2D-Ising universality class.
In Fig.~\ref{fig:sfac}b, the structure factor close to criticality is shown for two different system sizes of $96^2$ and $512^2$ lattice sites and the same set of simulation parameters. It is seen that, while the critical power-law behavior is well obtained in both cases in the intermediate wavenumber range, the structure factor shows quite pronounced finite-size effects at small $k$. Specifically, for the larger system, the Ornstein-Zernike type ``shoulder'' at low $k$ indicates that the system is still slightly above its critical point, whereas the low-$k$ excess seen for the smaller system apparently suggests that the system is already sub-critical.
However, according to the results of a FSS analysis performed on the order parameter and susceptibility (see below), not only the lager, but also the smaller system is still above its critical point.
Although finite-size effects appear to be more pronounced in the present case as compared to, for instance, molecular dynamics simulations \cite{hamanaka_onuki_pre2005}, it is well known that different quantities (e.g., the structure factor and the order-parameter distribution) in general show a different FSS behavior and are governed by their own apparent critical points \cite{toral_prb1990, amit_book}.
Here, for all studied parameter combinations and system sizes it is found that thermodynamic quantities are critical when the structure factor already displays slight effects of phase-separation (i.e., has an excess at low $k$).
It is clear from Fig.~\ref{fig:sfac}b that these discrepancies can in principle be reduced by using larger systems. However, the convergence appears to be quite slow in the present case.
Simulations performed deeper in the Ising regime furthermore suggest that this behavior is not directly associated with the proximity to the displacive limit.

\begin{figure*}[t]
\centering
    (a)\includegraphics[width=0.35\linewidth]{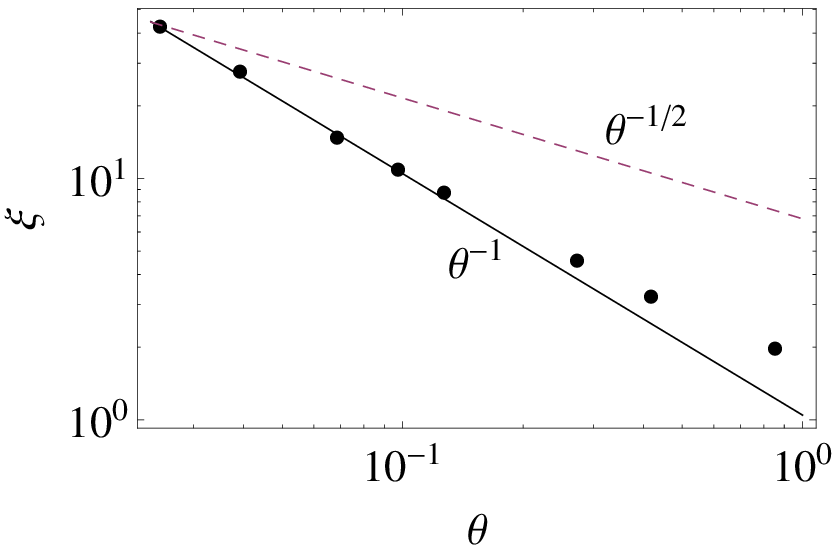}
    (b)\includegraphics[width=0.35\linewidth]{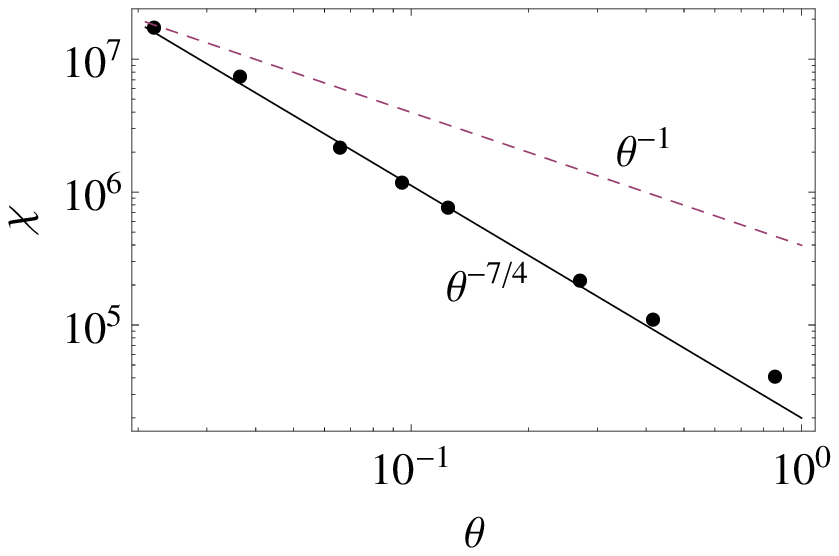}
    \caption{Critical growth of (a) the correlation length $\xi$ and (b) the susceptibility in dependence of the reduced temperature $\theta$ (for $\theta>0$). $\xi$ and $\chi$ are extracted from fitting Ornstein-Zernike forms [eq.~\eqref{CMF}] to the correlation function $C(k)$ [eq.~\eqref{G-comput}] obtained from simulations. Solid lines represent power-laws with 2D Ising exponents, while dashed lines are power-laws with mean-field exponents.}
    \label{fig:oz-fits}
\end{figure*}
The fact that sufficiently far above the critical point the structure factor assumes an Ornstein-Zernike form allows one to extract the critical growth of the correlation length $\xi$ and compressibility $\chi$ by fitting expression \eqref{CMF} to the simulation data for $C(k)$.
As Fig.~\ref{fig:oz-fits} shows, the correlation length and the compressibility approach the critical point by power-laws, $\xi \propto \theta^{-\nu}$ and $\chi\propto \theta^{-\gamma}$, with exponents that asymptotically agree with 2D-Ising values $\nu=1$ and $\gamma=7/4$.
Further away from the critical point ($\theta\gtrsim 0.2$), we observe cross-over to mean-field behavior.
It should be remarked that, especially in the case of the correlation length, the data admits in fact a certain range of fit values for the exponent $\nu$ and critical temperature $r_c$. For instance, in the present case it is found that the correlation length can be equally well described by an exponent of $\nu\approx 0.8$ and a slightly different $r_c$. Similar ``effective exponents'' have also been reported in previous Monte-Carlo simulations of the $\phi^4$-model \cite{milchev_1986} and reflect the fact that the width of the asymptotic region, where Ising-type behavior is observed, depends on the proximity to the displacive limit \cite{amit_book}. Indeed, approaching a critical point that is located closer to the ``Ising-limit'' on the critical line is found to already restrict the possible fit values for $\nu$ to a narrower margin around 1. Due to the finite size of the simulation box, however, it is not possible to follow the correlation length up to arbitrarily small reduced temperatures $\theta$.

\begin{figure}[t]
\centering
    \includegraphics[width=0.75\linewidth]{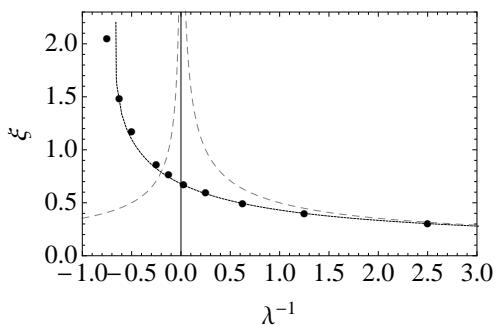}
    \caption{Dependence of the correlation length on the dimensionless coupling constant $\lambda^{-1}=r\kappa/u k_B T$ in the crossover regime from mean-field ($|\lambda|\ll 1$) to critical behavior ($\lambda\approx -1$). The dashed curve represents the mean-field correlation length, while the solid curve shows the prediction of perturbation theory, obtained from the numerical solution of eq.~\eqref{r-selfcon}. The symbols represent the correlation length extracted from Ornstein-Zernike fits to the structure factor obtained from simulations.}
    \label{fig:correl-crossover}
\end{figure}
In the crossover regime from mean-field to critical behavior, it is interesting to compare the simulation results with the predictions of perturbation theory (section \ref{sec:perturb}).
Figure~\ref{fig:correl-crossover} shows the correlation length as extracted from Ornstein-Zernike fits to the structure factor versus the inverse of the dimensionless coupling constant $\lambda^{-1} = r\kappa/u k_B T$, varying here only $r$. We see that, for $|\lambda^{-1}|\gg 1$, non-linear effects are negligible and the correlation length closely follows the mean-field prediction $\xi=(\kappa/r)^{1/2}$ (dashed line). Once $\lambda$ becomes of the order of unity, fluctuation corrections to mean-field behavior grow, leading to a suppression of the correlation length from its mean-field value (which diverges at $\lambda^{-1}=0$).
The solid curve in Fig.~\ref{fig:correl-crossover} represents the prediction for the renormalized correlation length $(\kappa/r')^{1/2}$ obtained from the numerical solution of the self-consistency equation \eqref{r-selfcon} for the parameter $r'$. We see that the simulation results for $\xi$ agree well with the theoretical predictions until $\lambda^{-1}\approx -0.6$.
The eventual breakdown of perturbation theory close to the critical point [i.e., for $\xi\gtrsim O(1)$] is of course expected, since self-energy contributions from all orders of the expansion diverge. Also, all wavenumber-dependent contributions to the renormalized parameters (which are strong in 2D) have been neglected here.
To increase the accuracy of the theoretical predictions in the 2D case, more sophisticated renormalization group methods would have to be employed \cite{kopietz_frg_book}.


\subsection{Finite-size behavior}
\begin{figure*}[t]
\centering
    (a)\includegraphics[width=0.3\linewidth]{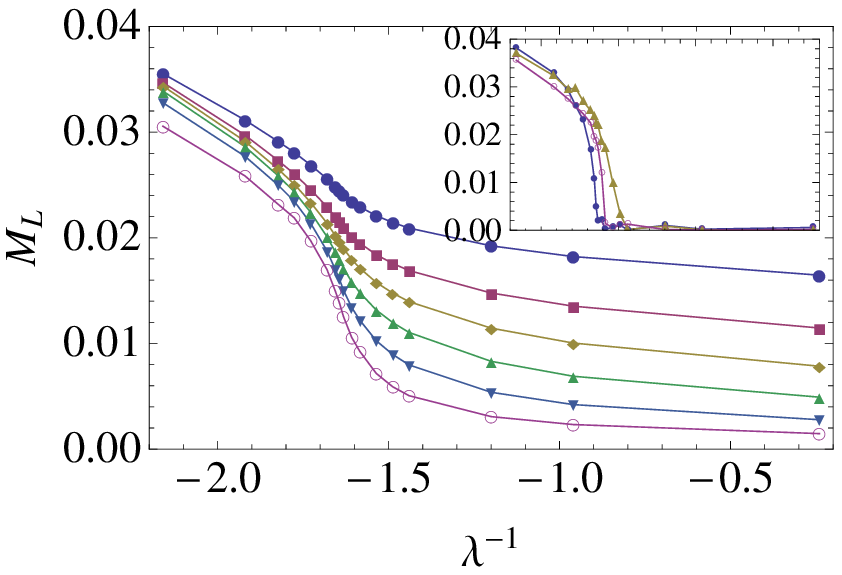}
    (b)\includegraphics[width=0.32\linewidth]{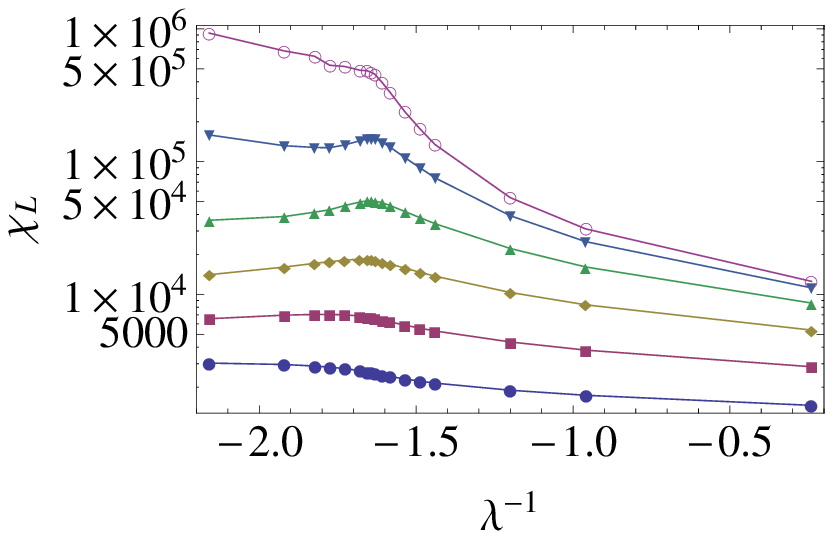}
    (c)\includegraphics[width=0.29\linewidth]{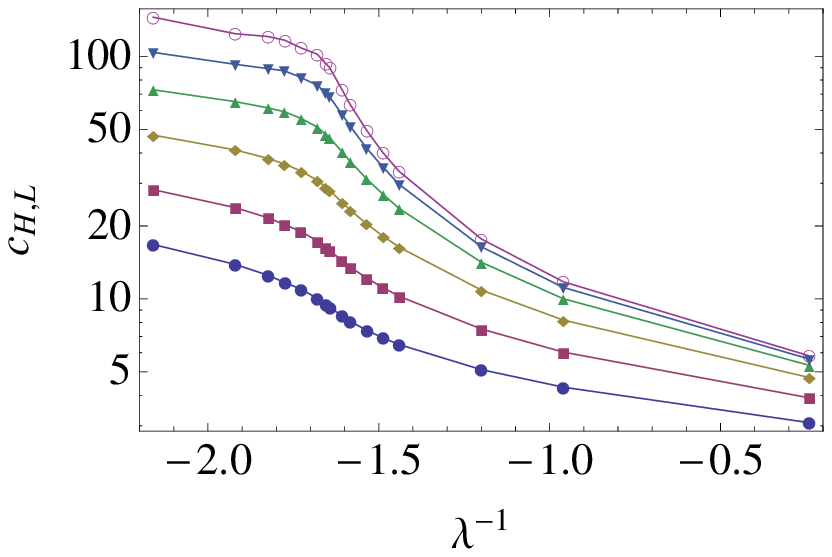}
    \caption{(Color online) Temperature dependence of (a) the coarse-grained order parameter $M_L$ [eq.~\eqref{ML}], (b) the susceptibility $\chi_L$ [eq.~\eqref{chiL}] and (c) the specific heat $c_{H,L}$ [eq.~\eqref{CL}] in the critical region. The inset in (a) shows the position of the peak of the order-parameter distribution $P_L$ (for visibility, only three curves are shown). The temperature is represented by the inverse dimensionless coupling $\lambda^{-1}\propto r$. The curves in each plot correspond to different subsystem dimensions $L$: $\bullet$ 1, {\tiny $\blacksquare$} 2, {\tiny $\blacklozenge$} 4, { $\blacktriangle$} 8, { $\blacktriangledown$} 16, {$\circ$} 32. Data for $L>S/8$ have been excluded. In all cases, $u=2.8\times 10^{-2}$, $\kappa=9.6\times 10^{-5}$, $k_B T=10^{-7}$.}
    \label{fig:temp}
\end{figure*}
Figure~\ref{fig:temp} shows the evolution of the block order parameter $M_L$, susceptibility $\chi_L$ and specific heat $c_{H,L}$ for different coarse-graining lengths $L$ in dependence of the temperature, which is represented here by the inverse dimensionless coupling $\lambda^{-1} = r\kappa/u k_B T$ ($\kappa$, $u$ and $T$ are fixed). The curves are drawn as a guide to the eye [they represent the FSS functions defined by eq.~\eqref{fss-ansatz}].
Passing from the high- to the low-temperature phase (i.e., decreasing $|\lambda^{-1}|$), the order parameter (Fig.~\ref{fig:temp}a) displays a sudden increase at around $\lambda^{-1}\approx -1.65$, which can be identified with the critical point. Due to the definition of $M_L$ as the average of the absolute value of the coarse-grained order parameter in each subbox [eq.~\eqref{ML}], $M_L$ is non-zero even in the disordered regime, but approaches zero with increasing subbox dimension $L$. Alternatively to eq.~\eqref{ML}, the order parameter can be defined by the position of the maximum of the underlying distribution $P_L$ (inset to Fig.~\ref{fig:temp}a), in which case the order parameter is exactly zero in the disordered phase (except in the immediate neighborhood of the critical point, where, in 2D, the order-parameter distribution develops a bimodal structure, cf.\ sec.~\ref{sec:op-dist}).
In the low-temperature phase, fluctuations decrease the average value of the order parameter with increasing coarse-graining length. Due to inevitable interfacial contributions in the coexistence regime, this effect is more pronounced for $M_L$ defined through eq.~\eqref{ML}.
For intermediate coarse-graining lengths $L$, the susceptibility -- which for simplicity is computed in Fig.~\ref{fig:temp}b via the same eq.~\eqref{chiL-lowT} in both the high- and low-temperature phase -- shows a peak at the apparent critical point, consistent with the behavior of $M_L$. For very small $L$, the peak is ``smeared out'' to a shoulder, while for the largest $L$ (not shown in the plot), the low-temperature data are strongly affected by interfacial contributions. Note that the peak positions are practically independent of $L$.
The specific heat (Fig.~\ref{fig:temp}c) shows a rapid increase around the critical point $\lambda\approx -1.65$, but no peak, in contrast to Monte Carlo simulations \cite{toral_prb1990, mehlig_zphysb1992}. The behavior of the specific heat seems to be similar to the order parameter $M_L$ and the absence of a peak might thus be related to the presence of interfacial contributions in the sub-critical regime.
It should be finally remarked here that the order parameter, susceptibility and specific heat obviously depend systematically on the coarse-graining length $L$. In order to obtain their true values, they have to be extrapolated to $L\ra\infty$. This does not affect the critical FSS behavior and is discussed further in sec.~\ref{sec:op-dist}.

\begin{figure*}[tb]
\centering
    (a)\includegraphics[width=0.3\linewidth]{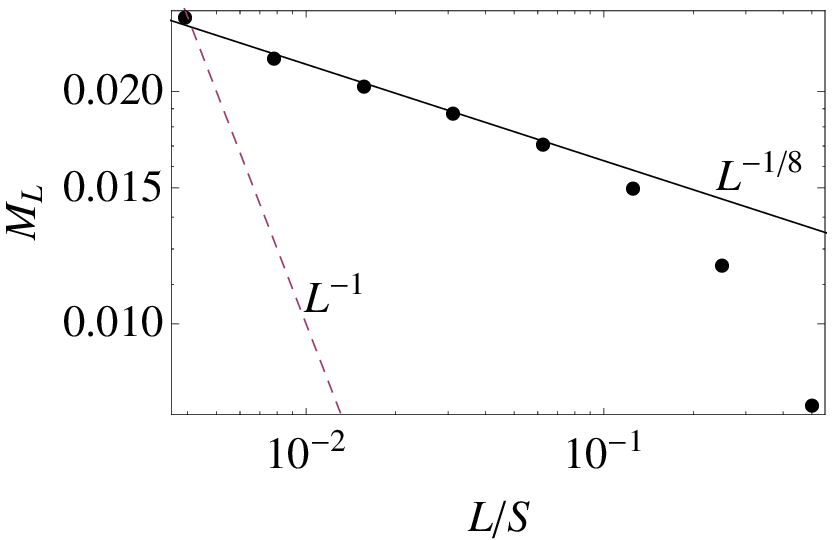}
    (b)\includegraphics[width=0.3\linewidth]{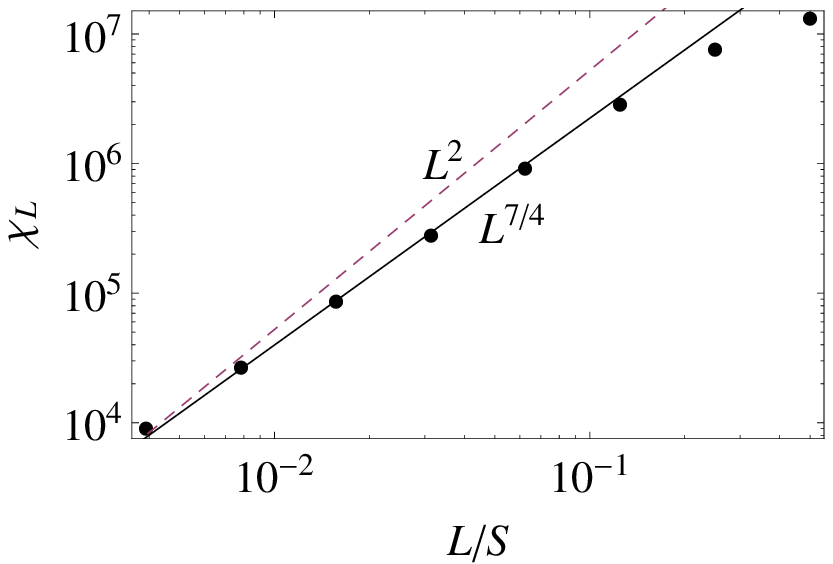}
    (c)\includegraphics[width=0.3\linewidth]{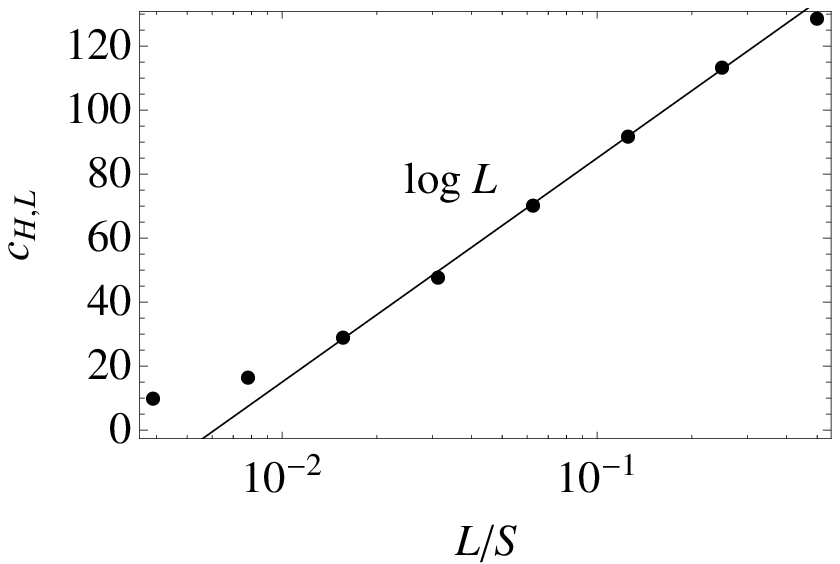}
    \caption{Finite-size scaling of (a) the block order parameter $M_L$, (b) the susceptibility $\chi_L$ and (c) specific heat $c_{H,L}$ with the block size $L$ (at fixed $S$) at the critical point. The solid lines represent the FSS predictions [eq.~\eqref{fss-crit2}] for the 2D-Ising case. For comparison, scaling laws according to mean-field theory are given by the dashed lines in (a) and (b). Simulation parameters are $r=-4.8\times 10^{-5}$, $u=2.8\times 10^{-2}$, $\kappa=9.6\times 10^{-5}$, $k_B T=10^{-7}$.}
    \label{fig:fss-crit}
\end{figure*}
In the immediate vicinity of the critical point, the theoretical correlation length exceeds the system size and one has to perform a FSS analysis to extract critical properties.
First, the scaling of the block order parameter $M_L$, susceptibility $\chi_L$ and specific heat $c_{H,L}$ with subbox-size $L$ at the critical point is investigated, keeping the lateral size $S$ of the simulation box (here, $S=256$) and all other simulation parameters fixed.
In this case, the FSS ansatz \eqref{fss-ansatz} predicts the scaling behavior
\beq \begin{split}
M_L &\sim L^{-\beta/\nu} g_M(L/S)\,,\\
\chi_L &\sim L^{\gamma/\nu} g_\chi(L/S)\,,\\
c_{H,L} &\sim \log(L) g_c(L/S)\,,
\end{split} \label{fss-crit2}\eeq
where $g_M$, $g_\chi$ and $g_c$ are scaling functions with limits $g_\Ocal(L/S)\rightarrow 0$ for $L\approx S$ due to the global order-parameter conservation and $g_\Ocal(L/S)\rightarrow \text{const.}$ for $L$ sufficiently smaller than $S$. Note that the temperature dependence has dropped out of the above scaling forms, as the temperature is kept fixed at its presumed critical value.
As Fig.~\ref{fig:fss-crit} shows, for $L\lesssim S/8$ the simulation results for $M_L$ and $\chi_L$ agree well with the FSS predictions of eq.~\eqref{fss-crit2} for the 2D-Ising case (solid lines in the plot). In case of the specific heat, the expected logarithmic scaling is obtained for $L\gtrsim 4$ and extends to block sizes up to $L\approx S/4$.
It is remarked that this close agreement can be obtained only in a rather narrow range around the critical point. Scaling plots like Fig.~\ref{fig:fss-crit} makes it possible, in principle, to estimate the true value of various intensive quantities by simple extrapolation of the straight line fits to the full system size $L=S$.

\begin{figure*}[t]
\centering
    (a)\includegraphics[width=0.38\linewidth]{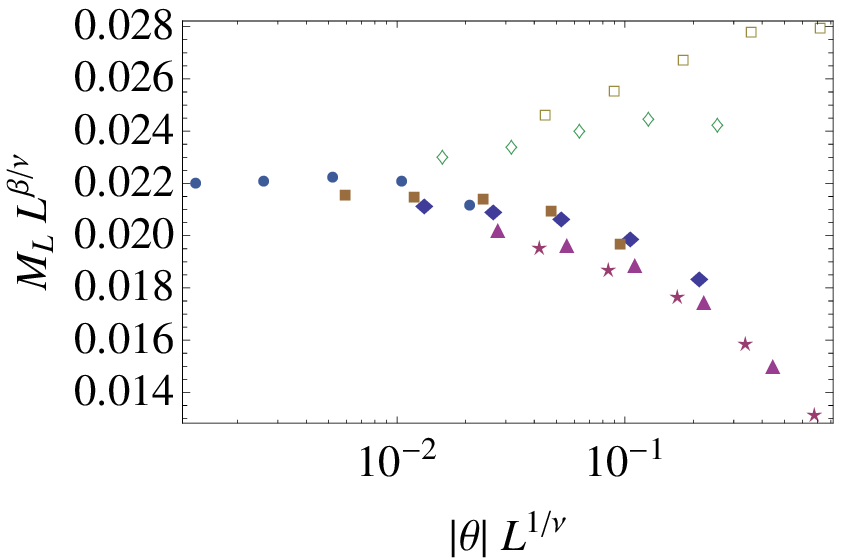}
    (b)\includegraphics[width=0.35\linewidth]{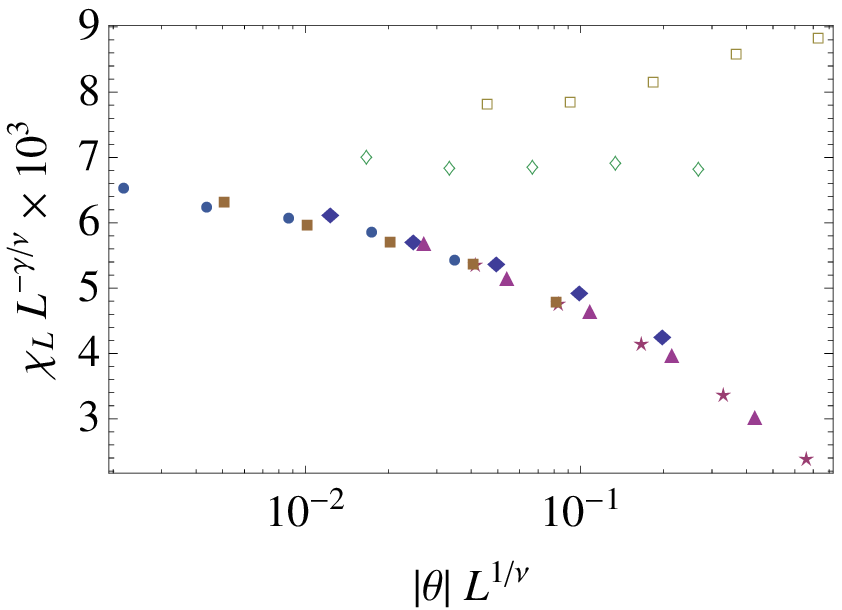}
    \caption{(Color online) Finite-size scaling plots of (a) the subbox order parameter $M_L$ and (b) susceptibility $\chi_L$. The exponents $\beta$, $\gamma$ and $\nu$ are fixed to 2D-Ising values. The critical temperature $r_c$ defining $\theta$ is set to $r_c=-4.82\times 10^{-5}$. Error bars represent an assumed uncertainity of $\Delta r-c/r_c\simeq 3\%$ in the critical temperature. Legend: $\star$ $\theta=0.042$, { $\blacktriangle$} $\theta=0.0028$, {\tiny $\blacklozenge$} $\theta=0.013$, {\tiny $\blacksquare$} $\theta=0.006$, $\bullet$ $\theta=-0.001$, {\tiny $\diamondsuit$} $\theta=-0.016$, {\tiny $\square$} $\theta=-0.044$. Data for $L=1$ and $L>S/8$ have been excluded.}
    \label{fig:fss-all}
\end{figure*}
Figure~\ref{fig:fss-all} shows the FSS behavior of the coarse-grained order parameter and susceptibility for varying subbox sizes $L$ \textit{and} temperatures $\theta$ in a combined plot. Data for $L=1$ as well as $L>S/8$ have been excluded, as they are expected to lie outside of the regime of validity of the FSS ansatz eq.~\eqref{fss-ansatz}.
In the plots, 2D-Ising exponents are used for the scaling transformation, together with a value for the critical temperature $r_c$ that is identical to the one obtained from the previous analysis.
The uncertainity in the value of $r_c$ represents a systematic error that affects the overall quality of the scaling behavior. The trends seen in Fig.~\ref{fig:fss-all} are found, however, to be quite insensitive to the specific value chosen for $r_c$.
The supercritical ($\theta>0$, solid symbols) branches of both the order parameter and the susceptibility show an acceptable scaling collapse. 
For the sub-critical branches (open symbols), however, neither the data for the order parameter nor for the susceptibility collapse onto a master curve. 
The main reason for the apparent scaling violation might be the global conservation of the order parameter, which leads to the coexistence of equal amounts of liquid and vapor below the critical point (for a quench at the critical density, which we consider here exclusively). The ensuing pronounced interfacial contributions to the order-parameter distribution (cf.~Fig.~\ref{fig:ophist-lowT}) might deteriorate scaling in the ordered phase. Similar effects have been pointed out in the context of lattice gas simulations in the canonical ensemble \cite{rovere_zphysb1993}\footnote{In standard grand-canonical Monte Carlo simulations, interfacial effects are much reduced as one stays in a pure phase most of the time \cite{binder_mc_intro}.}.
Additional influences on the scaling behavior can also arise from the fact that the $\phi^4$-model is equivalent to the Ising model only asymptotically close to the critical point and the width of the asymptotic region gets smaller with decreasing distance to the displacive limit ($r\rightarrow 0$, $u\rightarrow 0$) \cite{milchev_1986, amit_book}.
In fact, it is well known that the FSS form \eqref{fss-ansatz} represents only the leading order term of the full FSS expression \cite{wegner_prb1972, amit_book}, with the leading correction-to-scaling term being given by $L^{-\omega}g_\Ocal(L^{1/\nu}\theta)$ ($\omega=4/3$ in 2D). Thus, corrections to scaling are necessarily always present in a simulation. If a higher level of accuracy is desired, one might seek for an optimized set of coupling constants in the free energy functional, for which the leading-order scaling correction due to the dominant irrelevant operator is absent \cite{amit_book, ballesteros_perf_1998, hasenbusch_improv_1999}.
However, the gain in using an improved set of parameters might be spoiled by the presence of the additional scaling variable $L/S$, which in turn requires a rather large size $S$ of the total simulation box.

\begin{figure*}[t]
\centering
    (a)\includegraphics[width=0.35\linewidth]{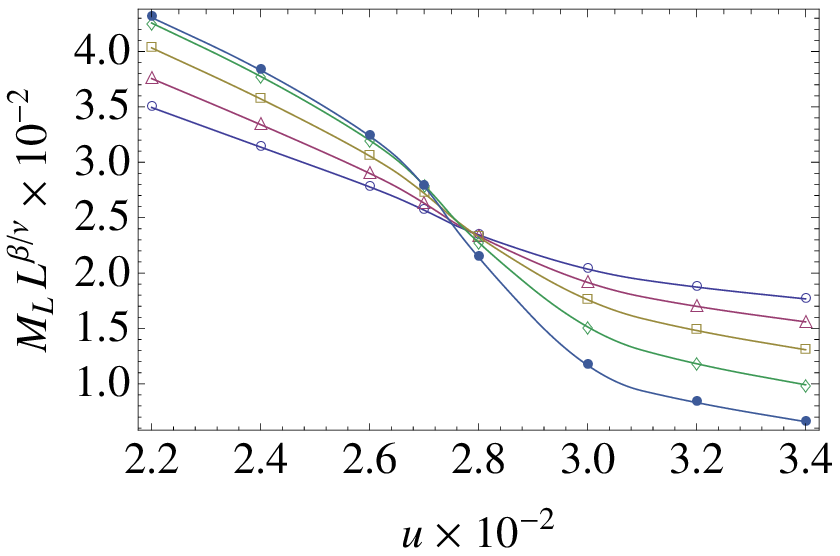}
    (b)\includegraphics[width=0.35\linewidth]{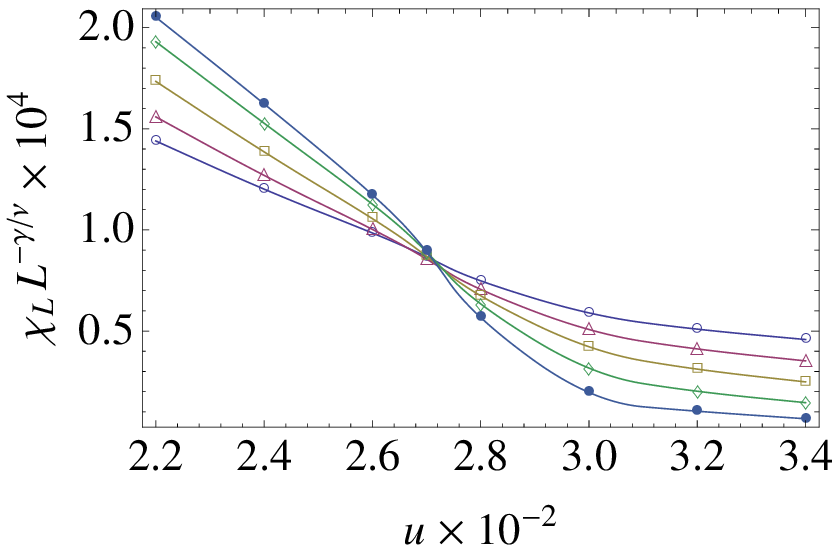}
    \caption{(Color online) Determination of the critical coupling by finite-size scaling analysis according to eq.~\eqref{fss-crit}. (a) Plot of the subsystem order parameter $M_L$ [eq.~\eqref{ML}] scaled by $L^{\beta/\nu}$ vs.\ the coupling $u$. (b) Plot of the subsystem compressibility $\chi_L$ [eq.~\eqref{chiL}] scaled by $L^{-\gamma/\nu}$ vs.\ the coupling $u$. In all cases, $r=-4.8\times 10^{-5}$ and $k_B T=10^{-7}$ and 2D-Ising values of the exponents $\beta$, $\gamma$ and $\nu$ are used. Different symbols correspond to subsystem sizes of $L=2$ ($\circ$), 4, ({\tiny $\triangle$}), 8 ({\tiny $\square$}), 16 ({\tiny $\diamondsuit$}), 32 ($\bullet$). Data for $L=1$ and $L>S/8$ are excluded from the plot. The curves are drawn as a guide for the eye.
    }
    \label{fig:fss-crit-coupl}
\end{figure*}
Finally, the usefulness of the FSS ansatz written in the form \eqref{fss-crit} to locate the critical point is demonstrated. In Fig.~\ref{fig:fss-crit-coupl}, the appropriately rescaled order parameter and susceptibility versus the non-linear coupling $u$ is plotted, keeping all other simulation parameters fixed. By eq.~\eqref{fss-crit}, the intersection point of all curves can be identified with the critical point $\theta=0$, which, for the present choice of simulation parameters, occurs for a value of $u\approx (2.7,\ldots, 2.8)\times 10^{-2}$. As expected, this value slightly depends on the quantity under consideration, but is otherwise consistent with the estimates of the critical point location from the FSS analysis of Fig.~\ref{fig:fss-crit}.


\subsection{Order-parameter distribution}
\label{sec:op-dist}
In the previous section, the FSS behavior of averaged thermodynamic quantities at the critical point was investigated primarily. We shall now turn to a more detailed study of the behavior of the underlying order-parameter probability distribution.

\begin{figure*}[t]
\centering
    (a)\includegraphics[width=0.36\linewidth]{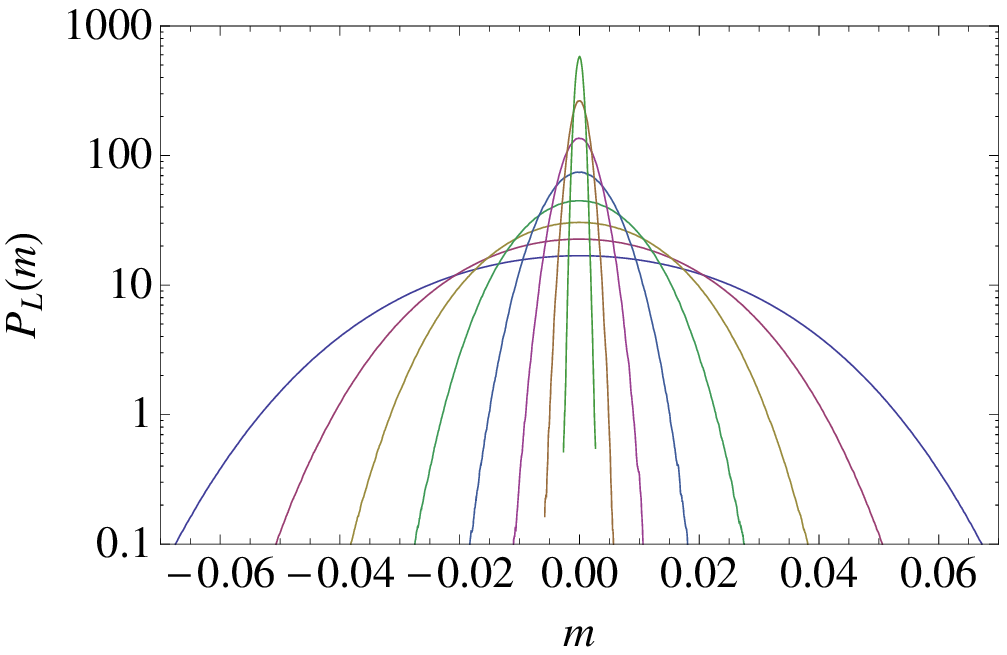}
    (b)\includegraphics[width=0.35\linewidth]{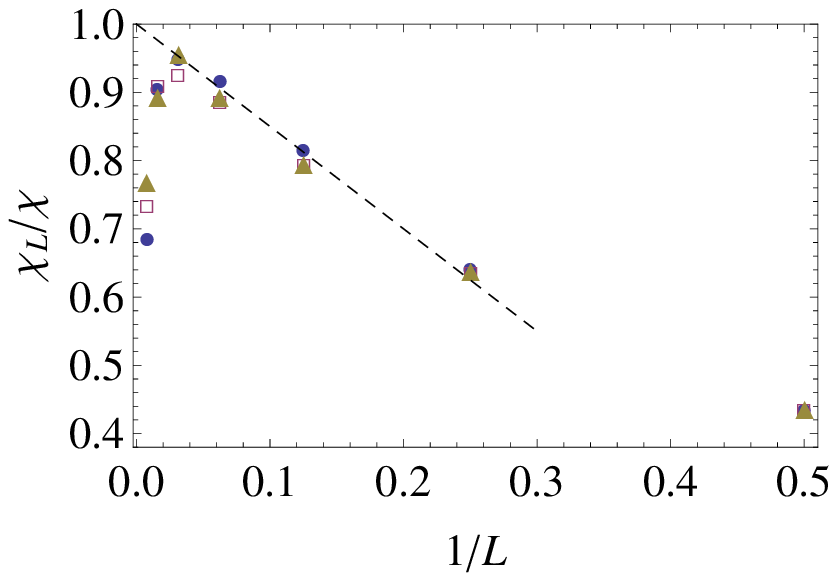}
    \caption{(Color online) Order-parameter distribution $P_L$ (a) and corresponding reduced susceptibility $\chi_L/\chi$ (b) for different $L$ far above the critical point (the correlation length is $\xi \simeq 1$). The individual curves in (a) correspond to different coarse-graining lengths $L$, where $L=1$ for the outer curve and $L=S/2$ for the innermost curve. For better visibility, individual data points are not shown. The subbox susceptibility $\chi_L$ in (b) is extracted from $P_L$ via Gaussian fits ($\bullet$), from the variance of $P_L$ ({\tiny $\square$}), and from the central height $P_L(0)$ ($\blacktriangle$). The expected susceptibility $\chi$ is computed from perturbation theory. The dashed line in (b) has a slope of $-1$.}
    \label{fig:ophist-highT}
\end{figure*}
Far above the critical point (Fig.~\ref{fig:ophist-highT}a), the order-parameter distribution has a perfectly Gaussian shape centered around the mean order-parameter value $m=0$. The variance decreases from the smallest block size ($L=1$) toward the largest ($L=S/2$), which is understandable from the fact that coarse-graining the system over a scale $L$ averages out fluctuations on smaller scales, which then do not contribute anymore to the variance.
As discussed in section \ref{sec:fss}, the coarse-grained susceptibility $\chi_L$, as determined by the width of $P_L$, in general differs from the true susceptibility obtained in the thermodynamic limit due to the neglect of correlations at the boundary of the subsystem. In particular, in the off-critical regime ($\xi<L$), $\chi_L$ is expected to differ from the true susceptibility by a correction factor $\sim 1/L$. This is demonstrated in Fig.~\ref{fig:ophist-highT}b, where $\chi_L$ is computed from $P_L$ by three different methods. As expected, the relation $\chi_L\sim 1/L$ holds in the range $\xi\ll L\ll S$, while for $L\sim S$, $\chi_L$ bends down toward zero due to the global order-parameter conservation. By extrapolating the linear part in $1/L$ toward $L\rightarrow \infty$, the true susceptibility can be estimated. Good agreement between the extrapolated value and the theoretical susceptibility is found.

\begin{figure}[t]
\centering
    \includegraphics[width=0.8\linewidth]{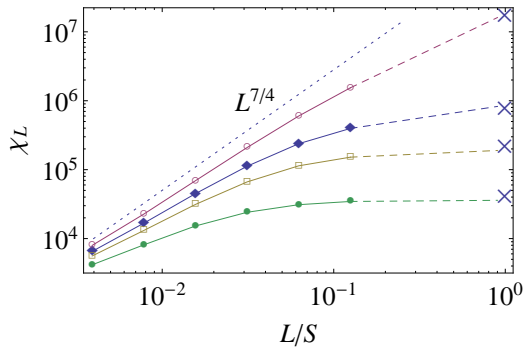}
    \caption{(Color online) System size dependence of the coarse-grained susceptibility $\chi_L$ obtained from the variance of the order-parameter distribution $P_L$ slightly above the critical point (where $L \lesssim \xi<S$). The dashed lines are extrapolations to the total system size $L=S$ and thus to the true susceptibility. The crosses represent the susceptibility obtained from the Ornstein-Zernike fits to the structure factor (cf.~Fig.~\ref{fig:oz-fits}). At the critical point, the susceptibility is expected to scale with the subbox size as $L^{\gamma/\nu}$ (dotted line) (cf.~Fig.~\ref{fig:fss-crit}b). The different symbols correspond to reduced temperatures of $\theta= 0.854$ ($\bullet$), 0.270 ({\tiny $\square$}), $0.124$ ({\tiny $\blacklozenge$}), 0.0219 ($\circ$). Data points for $L>S/8$ are excluded from the plot. The lines are drawn as a guide for the eye.}
    \label{fig:susc-precrit}
\end{figure}
In the critical regime ($\xi>L$), the corrections due to missing boundary correlations will clearly not be given anymore by a simple surface-to-volume ratio as above. Instead, the susceptibility will gradually approach its FSS form $\chi_L \sim L^{\gamma/\nu}$, as can be seen in Fig.~\ref{fig:susc-precrit}, where $\chi_L$ obtained from the variance of $P_L$ slightly above the critical point is plotted against $L$ (cf.\ Fig.~\ref{fig:fss-crit}). In the figure, the susceptibility data for $L>S/8$ is neglected due to a possibly spurious influence caused by the large value of $L/S$, for which the curves start to bend toward zero. Plotted in this way, one finds that extrapolating $\chi_L$ to $L\rightarrow S$ agrees well with the susceptibility obtained from Ornstein-Zernike fits to the structure factor of the entire system (Fig.~\ref{fig:oz-fits}).

\begin{figure*}[t]
\centering
    (a)\includegraphics[width=0.4\linewidth]{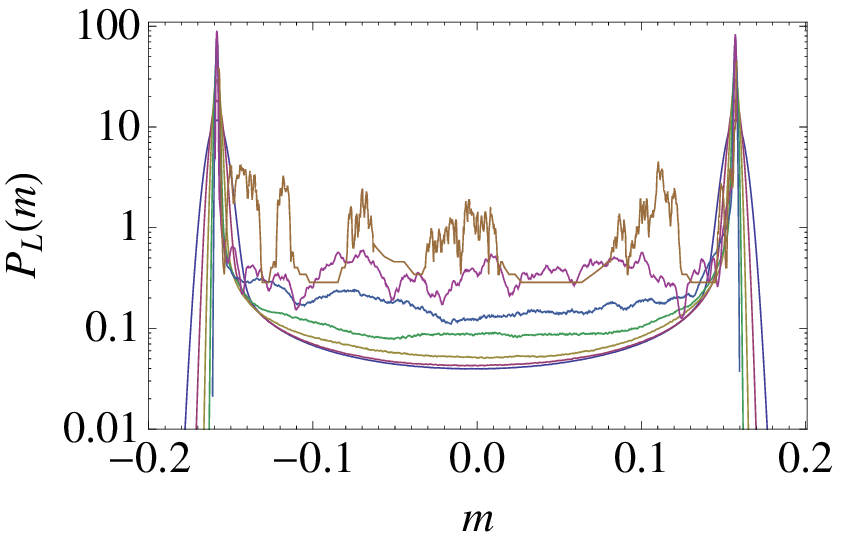}
    (b)\includegraphics[width=0.26\linewidth]{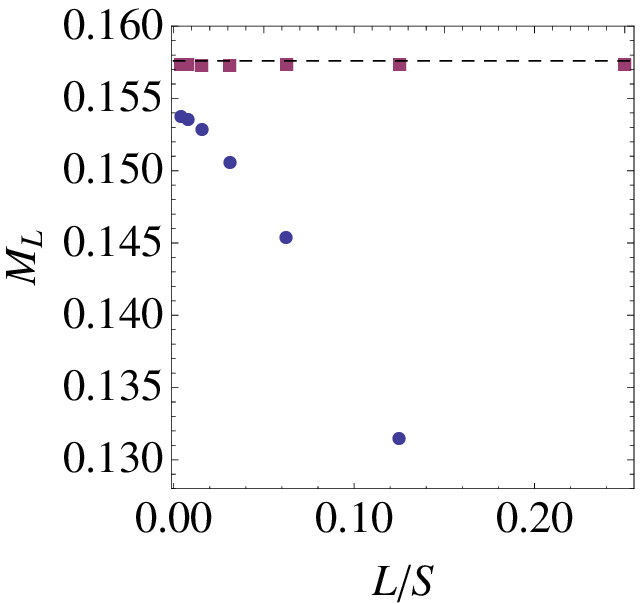}
    (c)\includegraphics[width=0.254\linewidth]{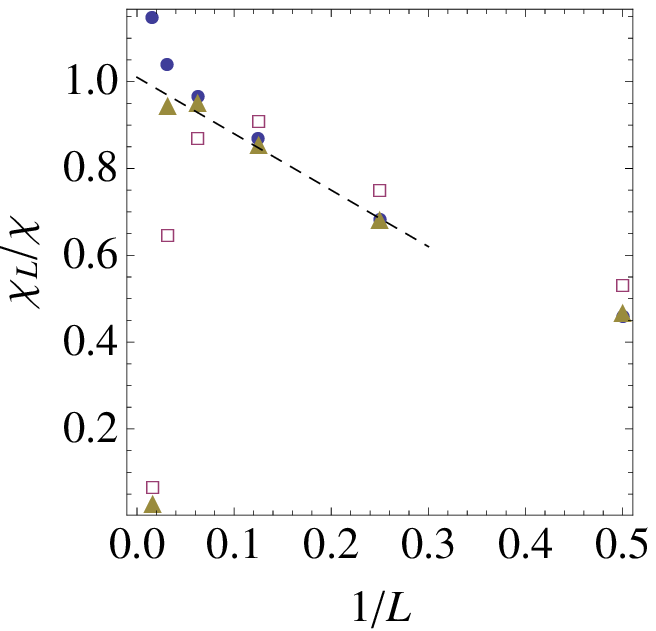}
    \caption{(Color online) Order-parameter distribution $P_L$ (a), order parameter $M_L$ (b) and susceptibility $\chi_L$ (c) for different coarse-graining lengths $L$ far below the critical point. The individual curves in (a) correspond to different $L$, where $L=1$ for the outer curve and $L=S/2$ for the innermost curve. In (b), the order parameter as defined by eq.~\eqref{ML} ($\bullet$) and by the peak position of $P_L$ ({\tiny $\blacksquare$}) is shown. The dashed line represents the value of $M_L$ expected from perturbation theory, eq.~\eqref{vev-implicit}. In (c), the susceptibility $\chi_L$ is extracted from $P_L$ via Gaussian fits ($\bullet$), from the variance of $P_L$ ({\tiny $\square$}), and from the height of the peaks $P_L(m\st{max})$ ($\blacktriangle$). The true susceptibility $\chi$ is computed from perturbation theory, eq.~\eqref{rsigma-broken}. The dashed line in (c) has a slope of $-1$.}
    \label{fig:ophist-lowT}
\end{figure*}
Distinctly below the critical point, the order-parameter distribution is characterized by two displaced Gaussians centered around the spontaneous order-parameter values $\pm \bra|m_L|\ket$ (Fig.~\ref{fig:ophist-lowT}a). Note that the probability distribution covers more than three orders of magnitude between its center and its peak.
The width of each Gaussian peak decreases with larger coarse-graining length $L$ as more and more fluctuations are averaged out.
The region between the peaks arises from interfacial configurations and is significantly in excess of a pure Gaussian contribution.
In agreement with the heuristic arguments outlined in section~\ref{sec:op-dist}, the central region does systematically increase with larger subsystem size $L$ and becomes approximately flat, as is expected by the presence of two-phase configurations with arbitrary proportions of liquid and vapor in each subbox.
However, even for the largest subbox sizes, the peaks are still far more dominant than the central region. This can be explained by two facts: First, interfacial free energies are still not negligible compared to bulk contributions, which would only be the case in the thermodynamic limit. Second, and more importantly, for deep quenches, the liquid domain (which in the present case is a single extended stripe) is not moving appreciably during the simulation time and thus each subbox will be mostly covered by the same, virtually static, phase configuration. This is also indicated by the strong irregularities found in the central region of the distributions for large $L$. To obtain the correct coarse-grained distribution, one would additionally have to perform an average over different simulation runs. This is, however, not attempted here.

\begin{figure*}[t]
\centering
   (a)\includegraphics[width=0.4\linewidth]{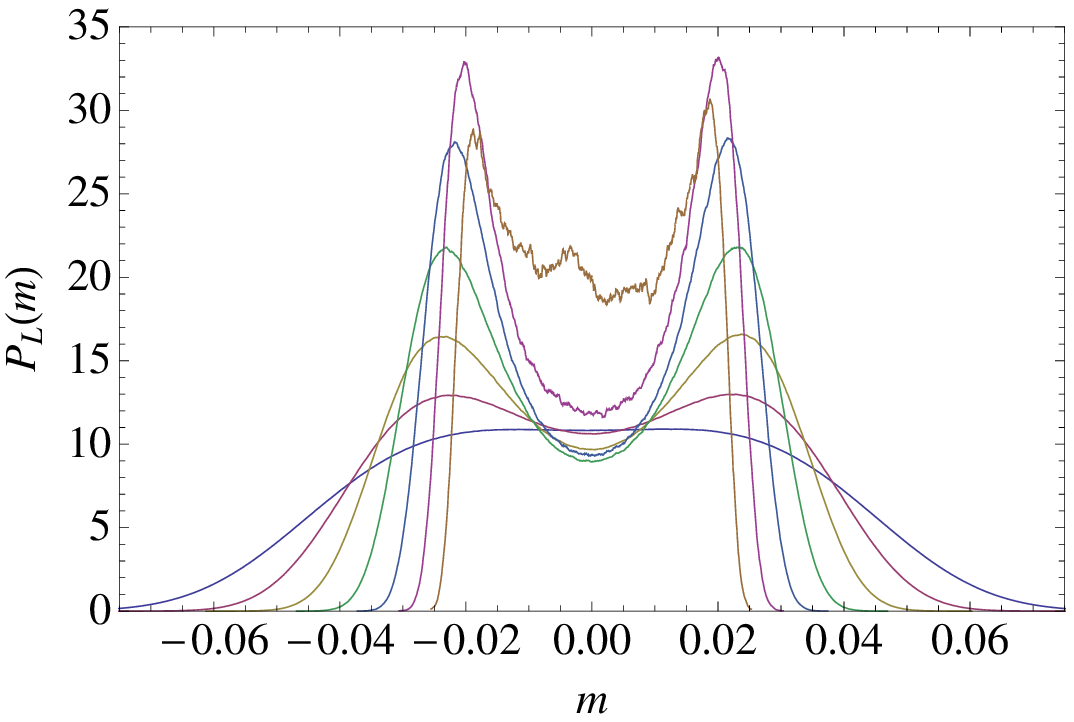}
   (b)\includegraphics[width=0.4\linewidth]{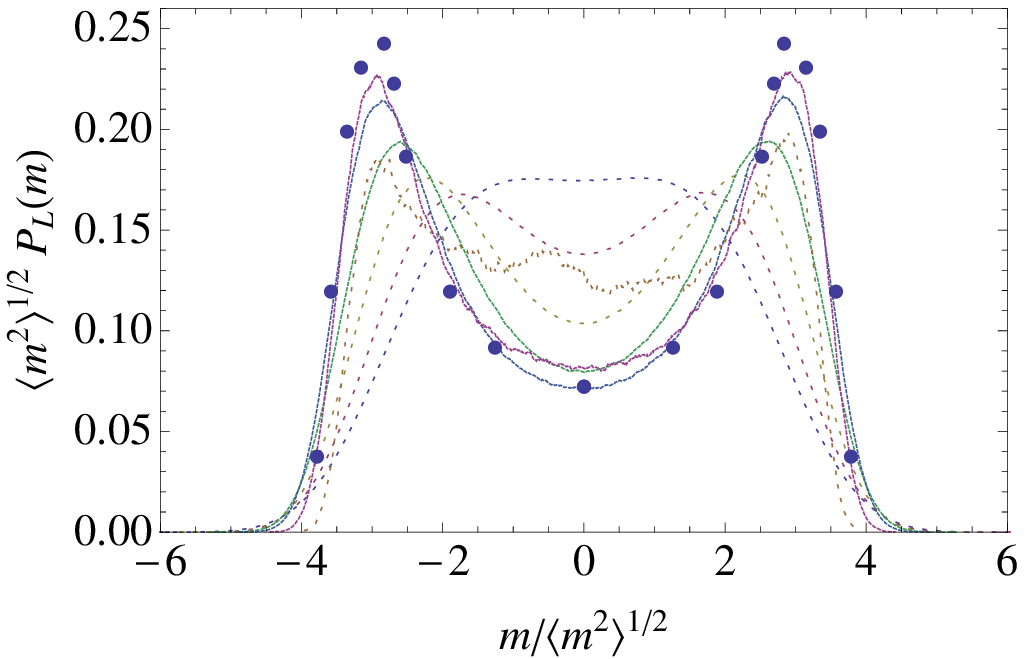}
    \caption{(Color online) Order-parameter distribution $P_L(m)$ for different $L$ at the critical point. In (a) the raw data is shown, while in (b) the distributions are rescaled by their standard deviation to compare with the predictions of eq.~\eqref{PL-fss}. For better visibility, curves in (b) that deviate from the scaling prediction are plotted with dashed lines. The solid circles in (b) represent the scaled distribution function from simulations of the 2D Ising model at the critical point by \cite{binder_zphysb1981}. The curves in each figure correspond to different coarse-graining lengths $L$, where in (a), $L=1$ for the outer curve and $L=S/2$ for the innermost curve; vice versa in (b). In all cases $S=256$. Note also the linear scale of the axes, in contrast to Figs.~\ref{fig:ophist-highT} and \ref{fig:ophist-lowT}.}
    \label{fig:ophist-crit}
\end{figure*}

In principle, the thermodynamic order parameter $M_L$ can be defined either by the average over half of the distribution, $M_L=\bra |m_L|\ket$ [eq.~\eqref{ML}], or by the position $m\st{max}$ of the maximum of $P_L(m)$.
In the coexistence regime, one finds that the latter definition is in general in closer agreement to the theoretical prediction, eq.~\eqref{vev-implicit}, for all values of $L$ (Fig.~\ref{fig:ophist-lowT}b). In principle, a slight system-size dependence is always expected as fluctuations in general tend to reduce the average order-parameter value, which is clearly seen closer to criticality (inset to Fig.~\ref{fig:temp}a and Fig.~\ref{fig:ophist-crit}a).
The order parameter defined by eq.~\eqref{ML} strongly decreases with larger $L$ due to contributions from phase-separated states to the average of $|m_L|$.
Thus, far above or below the critical point, defining the order parameter $M_L$ as the location of the maximum of $P_L$ seems in general preferable over the definition of eq.~\eqref{ML}, since the former ensures that $M_L$ is exactly zero in the high-temperature phase and has a negligible dependence on the system size or on interfacial contributions in the low-temperature phase. In contrast, the definition of eq.~\eqref{ML} behaves smoother in the critical region and is therefore better suited for FSS analyses.

Analogously to the high-temperature case, the coarse-grained susceptibility can be obtained either from the peak height [$\chi_L \simeq L^d / 8\pi k_B T P_L^2(m\st{max})$, see eq.~\eqref{PL-lowT}] or the peak variance. As all these methods implicitly assume the presence of two displaced Gaussians [eq.~\eqref{PL-lowT}], the central region of the distribution should be excluded beforehand\footnote{In particular, the normalization should be computed with the central region set to zero}.
Due to the significant asymmetry of the wings and the pronounced interfacial contributions, the variance is thus most reliably obtained by fitting Gaussians to the peaks.
It is seen from Fig.~\ref{fig:ophist-lowT}c, that all the three different estimates of the susceptibility roughly agree, except for values of $L$ close to the total system size.
Extrapolating the linear part in $1/L$ of the coarse-grained susceptibilities $\chi_L$ to the limit $L\rightarrow \infty$ makes it possible to obtain the true susceptibility.

\begin{figure*}[t]
\centering
   (a)\includegraphics[width=0.36\linewidth]{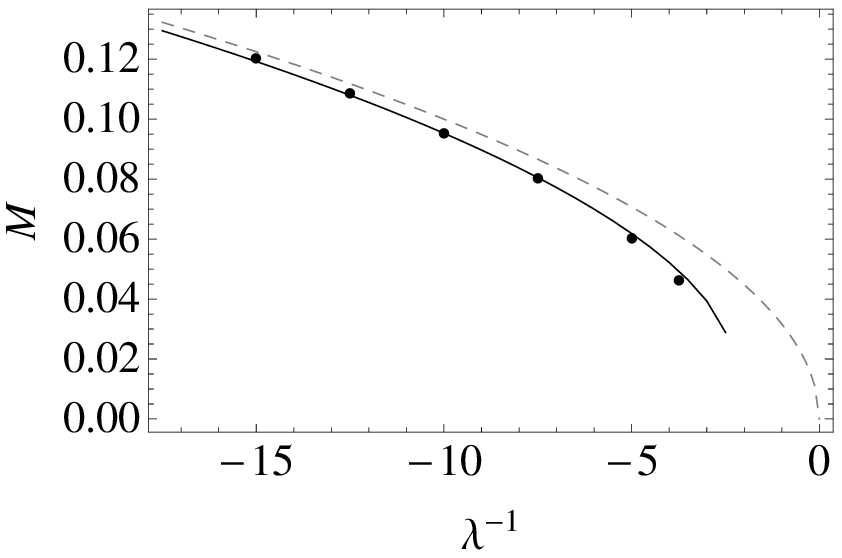}
   (b)\includegraphics[width=0.38\linewidth]{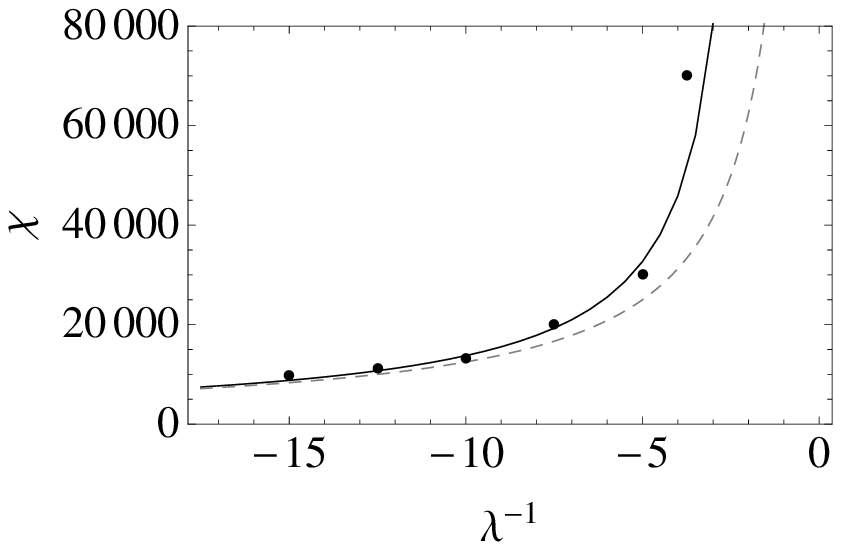}
    \caption{Temperature dependence of (a) the thermodynamic order parameter $M$ and (b) the susceptibility $\chi$ in the phase-coexistence regime. The temperature is given here in terms of the inverse dimensionless coupling $\lambda^{-1}\propto r$. The symbols ($\bullet$) represent simulation results, the solid curve represents the predictions of perturbation theory [eqs.~\eqref{vev-implicit} and \eqref{rsigma-broken}] and the dashed curve represents the prediction of mean-field theory. The critical point is located here at $\lambda^{-1}\approx -1.5$.}
    \label{fig:eos-lowT}
\end{figure*}
In Fig.~\ref{fig:eos-lowT}, the order parameter and susceptibility in dependence of the temperature (represented by the inverse dimensionless coupling $\lambda^{-1}\propto r$) are compared to the predictions of perturbation theory (see sec.~\ref{sec:perturb}). The order-parameter data shown in Fig.~\ref{fig:eos-lowT}a are obtained from the location of the peak of the distribution, which is roughly independent of $L$ and has negligible statistical scatter (see Fig.~\ref{fig:ophist-lowT}b). One sees that the simulation results for the order parameter agree well with the prediction of eq.~\eqref{vev-implicit} for $\bra \phi\ket$ including the leading-order fluctuation corrections. As expected, deviations become noticeably closer to the critical point (here, $\lambda^{-1}\approx -1.5$), where a perturbative treatment is not applicable. One further notes that fluctuations generally tend to reduce the order-parameter below its mean-field value, even relatively far away from the critical point.
The data for the susceptibility shown in Fig.~\ref{fig:eos-lowT}b is obtained by extrapolating the block susceptibility $\chi_L$ found from the peak height of the distribution to $L\ra\infty$ (cf.\ Fig.~\ref{fig:ophist-lowT}c).
In contrast to the order parameter, the susceptibility data exhibits significantly stronger statistical scatter -- in particular, closer to criticality.
Nevertheless, for the temperature range investigated, acceptable agreement between simulation results and the predictions of perturbation theory for $\chi$ in the symmetry-broken phase, eq.~\eqref{rsigma-broken}, is found.


In Fig.~\ref{fig:ophist-crit}, the coarse-grained order-parameter distribution at the critical point is shown. Interestingly, for sufficiently large coarse-graining lengths, the distribution shows a pronounced double-peak structure, which is found to persist even slightly above the critical point. This is in agreement with Monte-Carlo results for the two-dimensional Ising model \cite{binder_zphysb1981} and renormalization group calculations \cite{bruce_jphysc1981}. For small coarse-graining lengths, where the distribution essentially probes non-universal properties, $P_L$ depends significantly on the location on the critical line: Close to the displacive limit, $P_1$ appears concave, whereas toward the Ising-limit, it develops a double-peak structure.
This is understandable since the gradient term in the free energy functional dominates over the bare Landau potential for high degrees of displaciveness.
The scaling ansatz \eqref{PL-fss} for the critical order-parameter distribution predicts that when expressing the data in terms of the scaled variables $m L^{\beta\nu}$ and $P_L L^{-\beta/\nu}$, all points should collapse on a single curve. In order to compare our results with the Ising model calculations of ref.~\cite{binder_zphysb1981}, the rescaling procedure is implemented here by appropriately multiplying the data by the standard deviation $\bra m^2\ket^{1/2}$, which is expected to be equivalent concerning the overall scaling behavior since $\bra m^2\ket \sim L^{-2\beta/\nu}$. 
As Fig.~\ref{fig:ophist-crit}b shows, the scaling of the distribution function predicted by eq.~\eqref{PL-fss} holds in a range $1\ll L\ll S$. This might seem surprising insofar, as the FSS of the low-order moments of $P_L$ (the coarse-grained order parameter and susceptibility) works well already for the smallest box sizes (see Fig.~\ref{fig:fss-crit}). However, the full probability distribution obviously contains more information than just its low-order moments, and thus a scaling of $P_L$ is only a sufficient, but not necessary condition for the scaling of $M_L$ and $\chi_L$. In fact, in the present case, the fourth-order cumulant already fails to show the well-known scaling behavior observed in standard (grand-canonical) Monte-Carlo simulations of the Ising-model \cite{binder_zphysb1981}. A similar behavior has also been observed in lattice gas simulations with a conserved order parameter \cite{rovere_zphysb1993}. The scaling behavior of the distribution for smaller coarse-graining lengths is found to improve with increasing distance from the displacive limit.
A direct comparison of $P_L$ in the scaling regime to the corresponding order-parameter distribution of the two-dimensional Ising model at criticality \cite{binder_zphysb1981}, represented by the solid points in Fig.~\ref{fig:ophist-crit}b \footnote{To match the width of the distributions used here, the Ising model data is rescaled as $(m,P_L)\rightarrow (m/c,c P_L)$ with a factor $c$ as allowed by the scaling ansatz.}, shows close agreement, except for a slight underestimation of the peak heights.

\section{Summary}
In this work, static critical phenomena of a one-component fluid have been studied using a fluctuating non-ideal gas LB model. In this model, the fluctuating Navier-Stokes equations for the density and momentum of a compressible, isothermal fluid based on a Ginzburg-Landau $\phi^4$-free energy functional are solved.
It is found that the model is able to capture the essential features of the static critical behavior associated with the 2D Ising universality class.
A characteristic property of the present simulation method is the global conservation of the order parameter, which demands a careful interpretation of FSS results.
The conserved nature of the order parameter leads to the presence of coexisting two-phase states below the critical point and is expected to be the main source of scaling corrections in the present case.
Despite these complications, the expected critical behavior of the structure factor, order parameter, susceptibility and specific heat is found to be overall well reproduced.
However, it was noted that finite-size effects appear to have a quite strong effect on the structure factor, which assumes its expected critical scaling law at a slightly higher temperature than the other thermodynamic observables.
For future work, it would be interesting to compare these results to other LB models of non-ideal fluids.
The order-parameter distribution function, which contains useful information on two-phase states below the critical point, compares well with theoretical predictions and Ising model calculations.
Also, issues relevant to coarse-graining and generic fluctuation induced effects on observable quantities near and far from the critical point have been discussed.

The present work has only dealt with static critical phenomena. While it is clear that Monte Carlo methods are usually better suited for this task, an assessment of the LB method in this regard is nevertheless important since the successful reproduction of equilibrium aspects is a necessary prerequisite for a faithful application of the method to, for instance, dynamical problems. Also, understanding the equilibrium behavior of the model and its coarse-graining properties is important for many practical problems employing an effective free energy description, such as nucleation and spinodal decomposition. The present work is thus hoped to provide a useful starting point for further applications of the LB method to problems of current interest involving phase-transitions and critical phenomena in fluids.

\vrule

\acknowledgements
We would like to thank R. Adhikari, K.\ Binder, M.\ E.\ Cates, J.\ Horbach, A.\ Troester and A.\ J.\ Wagner for useful discussions.
Funding from the industrial sponsors of ICAMS, the state of North-Rhine Westphalia and the European Commission in the framework of the European Regional Development Fund (ERDF) is gratefully acknowledged.

\appendix


\end{document}